\documentclass[usenatbib,usegraphicx,twocolumn]{mn2e}
\usepackage[english]{babel} 
\usepackage{color}
\usepackage{xcolor}
\usepackage{bigfoot}
\usepackage{ulem} % \sout{Bill Clinton}
\usepackage{url}
\usepackage{hyperref}

\def\u{\vec{U}}

\def\araa{ARAA}

\def\na{New Astronomy}
\def\pasp{PASP}
\def\mnras{MNRAS}

\def\aap{A \& A}
\def\aaps{Astronomy and Astrophysics Supplement Series}
\def\apj{ApJ}

\def\u{{\bf U}} 

\def\V2{V_2}
\def\V2ij{V_{2ij}}
\def\S{{\mathcal S}}
\def\V{\mathcal{V}}

\def\la{\left\langle}

\def\lsim{~\rlap{$<$}{\lower 1.0ex\hbox{$\sim$}}}

\def\gsim{~\rlap{$>$}{\lower 1.0ex\hbox{$\sim$}}}

\usepackage{graphics}
\usepackage{color}
\usepackage{amsmath}
\usepackage{amssymb}
\usepackage{psfrag}
\usepackage{subfig}
\usepackage{graphicx}

\newsavebox{\measurebox}
\begin{document}
\date {} 
\title[Angular power spectrum using TGSS survey] {All sky angular power spectrum: I. Estimating brightness temperature fluctuations using TGSS 150 MHz survey}
\author[S. Choudhuri et al.]{Samir Choudhuri$^{1,2}$\thanks{Email:s.choudhuri@qmul.ac.uk}, Abhik Ghosh$^{3,4,5}$, Nirupam Roy$^{6}$, Somnath Bharadwaj$^{7}$,
\newauthor Huib.~T.~Intema$^{8}$ and Sk. Saiyad Ali$^{9}$\\
$^{1}$ Astronomy Unit, Queen Mary University of London, Mile End Road, London E1 4NS, United Kingdom\\
$^{2}$ National Centre For Radio Astrophysics, Post Bag 3, Ganeshkhind, Pune 411 007, India\\
$^{3}$ Department of Physics, Banwarilal Bhalotia College, GT Rd, Ushagram, Asansol, West Bengal, India \\
$^{4}$ Department of Physics and Astronomy, University of the Western Cape, Robert Sobukwe Road, Bellville 7535, South Africa\\
$^{5}$ The South African Radio Astronomy Observatory (SARAO), 2 Fir Street, Black River Park, Observatory, 7925, South Africa\\
$^{6}$ Department of Physics, Indian Institute of Science, Bangalore 560012, India\\
$^{7}$ Department of Physics,  \& Centre for Theoretical Studies, IIT Kharagpur,  Kharagpur 721 302, India\\
$^{8}$ Leiden Observatory, Leiden University, Niels Bohrweg 2, NL-2333CA, Leiden, The Netherlands\\
$^{9}$ Department of Physics,Jadavpur University, Kolkata 700032, India\\}
\maketitle

\begin{abstract}
Measurements of the Galactic synchrotron emission is relevant for the 21-cm studies from the Epoch of Reionization. The study of the synchrotron emission is also useful to quantify the fluctuations in the magnetic field and the cosmic ray electron density of the turbulent interstellar medium (ISM) of our Galaxy. Here, we present the all-sky angular power spectrum $(C_{\ell})$ measurements of the diffuse synchrotron emission using the TIFR GMRT Sky Survey (TGSS) at 150 {\rm MHz}. We estimate $C_{\ell}$ using visibility data both before and after subtracting the modelled point sources. The amplitude of the measured $C_{\ell}$ falls significantly after subtracting the point sources, and it is also slightly higher in the Galactic plane for the residual data. The residual $C_{\ell}$ is most likely to be dominated by the Galactic synchrotron emission. The amplitude of the residual $C_{\ell}$ falls significantly away from the Galactic plane. We find the measurements are quite symmetric in the Northern and Southern hemispheres except in the latitude range $15-30^{\circ}$ which is the transition region from the disk dominated to diffuse halo dominated region. The comparison between this interferometric measurement with the scaled version of the Haslam rms map at 150 {\rm MHz} shows that the correlation coefficient $(r)$ is more than 0.5 for most of the latitude ranges considered here. This signifies the TGSS survey is quite sensitive to the diffuse Galactic synchrotron radiation.
\end{abstract} 


\section{Introduction}
\label{intro}
The redshifted 21-cm signal from neutral hydrogen (HI) has been perceived to be one of the most promising probes of the epoch of reionization (EoR) (see \citealt{furlanetto06,morales10,pritchard12,mellema13} for reviews).  The hydrogen in the universe changes its phases from the neutral to the ionized state in this epoch, and many issues like the exact time and duration of reionization, and the sources responsible for this process are still unresolved. Several ongoing and future radio telescopes such as the Low Frequency Array
(LOFAR{\footnote{\url{http://www.lofar.org/}}}, \citealt{haarlem}), the Murchison Wide-field Array (MWA{\footnote{\url{http://www.mwatelescope.org}}} \citealt{bowman13}), the Square Kilometer Array (SKA1 LOW{\footnote{\url{http://www.skatelescope.org/}}}, \citealt{koopmans15}) and the Hydrogen Epoch of Reionization Array (HERA{\footnote{\url{http://reionization.org/}}}, \citealt{deboer17}) including existing the Giant Metrewave Radio Telescope (GMRT \footnote{\url{http://www.gmrt.ncra.tifr.res.in}}; \citealt{swarup,paciga13}) are seeking to measure the 21-cm signal from the EoR.

The presence of strong astrophysical foregrounds that are 4-5 orders of magnitude brighter than the expected 21-cm signal \citep{shaver99,dmat1,santos05,ali,paciga11,ghosh1} poses a big challenge for the detection of the EoR 21-cm signal. The major foreground components include the extra-galactic radio point sources, the diffuse
Galactic synchrotron emission (DGSE), and Galactic and extra-galactic free-free emission. The extra-galactic point sources are the most dominant foreground components at that angular scales which are relevant for telescopes like LOFAR and SKA \citep{ali,ghosh150}. The DGSE dominates at large angular scale $ >10$ arcmin after point sources are subtracted at $\sim$10-20 mJy level \citep{bernardi09,ghosh150,iacobelli13,samir17a}.

The DGSE is produced by the cosmic ray electrons spiralling in the Galactic magnetic field lines \citep{ginz69,rybicki79}. A precise characterization and a detailed understanding of the DGSE are needed to remove foregrounds in 21-cm experiments reliably. Also, the angular fluctuations of the DGSE are directly related to the fluctuations in the magnetic field and the fluctuations in the cosmic ray electron density of the turbulent interstellar medium (ISM) of our Galaxy \citep{cho08,Waelkens,regis11,Lazarian,iacobelli13}, a subject that is not very well understood at present. \citet{Lazarian} have shown avenues for quantitative studies of magnetic turbulence in our Galaxy and beyond using observations of the synchrotron emission, and it also outlined the directions of how synchrotron foreground emission can be separated from the cosmological signal, i.e., from cosmic microwave background or highly redshifted HI 21-cm emission.

Several observations spanning  a wide range of frequencies have characterized  different aspects of the DGSE \citep{haslam81,haslam82,reich82,reich88,jonas98,ellin13}. \citet{guzz11} have produced an all-sky spectral index map of the DGSE between 45 and 408 {\rm MHz} using their own all-sky map at 45 {\rm MHz}, the 45 {\rm MHz} southern and northern sky maps \citep{alvarez97,maeda99} and the 408 {\rm MHz} all-sky map \citep{haslam81,haslam82}. The Global Sky Model (hereafter, GSM) for the diffuse Galactic emission temperature map has been developed in the frequency range 10 {\rm MHz} to 94 {\rm GHz} based on 11 most accurate data sets using principal component analysis \citep{costa08}. \citet{zheng17} have produced an improved GSM of the diffuse Galactic radio emission from 10 {\rm MHz} to 5 {\rm  THz} which includes 29 sky maps. These type of models are highly useful to understand the Galactic foreground contributions in wide-band CMB and cosmological 21 cm HI observations.

The statistical properties of the DGSE can be quantified in terms of the angular power spectrum $C_{\ell}$. Various authors have used the above mentioned all-sky observations to estimate the statistical properties of the DGSE for a wide range of frequencies \citep{tegmark96,bouchet99,giardino01,giardino02,bennett03}. The $C_{\ell}$ of the DGSE intensity fluctuations spanning over large portions of the sky can be modelled by a power law i.e. $C_{\ell}\propto\ell^{-\beta}$ \citep{tegmark00,bacci01}. \citet{laporta08} have analysed the 408-{\rm MHz} Haslam map \citep{haslam81,haslam82} and the 1420-{\rm MHz} survey data \citep{reich82,reich86,reich01} separately to measure the $C_{\ell}$ of the DGSE and found $\beta$ values in the range 2.6 - 3.0 down to the angular multipoles  of $\ell = 200$ and 300 at 408 and 1420 {\rm MHz} respectively. These studies show that  $\beta$ steepens (or increases)  towards higher Galactic latitude.

The properties of the angular power spectrum of the DGSE are not well quantified at the frequencies and angular scales relevant for
detecting the cosmological 21-cm signal from the EoR. \citet{parsons10} have presented the all-sky synthesized map and estimated the $C_{\ell}$ between 139 {\rm MHz} and 174 {\rm MHz}. It has also been measured in only a few small fields at low Galactic latitude $|b| <
14^{\circ}$ in the frequency range 150 - 160 {\rm MHz} \citep{bernardi09,bernardi10,ghosh150,iacobelli13,samir17a}. \citet{bernardi09} and \citet{ghosh150} have, respectively, analysed 150 {\rm MHz} WSRT and GMRT observations where they respectively found $\beta=2.2 \pm 0.3$ and $\beta=2.34 \pm 0.28$ up to $\ell=900$. \citet{iacobelli13} have measured the $C_{\ell}$ of the DGSE at 160 {\rm MHz} using LOFAR data and reported that the angular power spectrum has a  slope $\beta \approx 1.8$ down to the angular multipoles $\ell$ of 1300. In an earlier paper \citep{samir17a}, we have analysed two fields from the TIFR GMRT Sky Survey (TGSS)-ADR1 survey at 150 {\rm MHz} \citep{intema17} and measured the $C_{\ell}$ of the DGSE across the $\ell$ range $240 \le \ell \le500$ and  found that the values of $\beta$ are $2.8 \pm 0.3$ and $2.2 \pm 0.4$ respectively in the two fields. Recenly \citet{chakra19} have measured the $C_{\ell}$ in ELAIS-N1 field and found the $\beta$ values consistent with earlier measurements. All of these results are restricted to small portion of the sky $\le 6^{\circ} \times 6^{\circ}$.

The GMRT field of view has a FWHM of $3.1^{\circ}${\footnote{\url{http://gmrt.ncra.tifr.res.in/}}} at $150 \, {\rm MHz}$. The TGSS \citep{sirothia14} contains observations in $5336$ pointings covering a large fraction ($90 \%$) of the total sky in the declination range $\delta >-55^{\circ}$. Here we have used the first alternative data release (ADR1) of the  TGSS  that was calibrated and processed by \citet{intema17}. In this paper, we have applied the visibility based Tapered Gridded Estimator (TGE) \citep{samir16b} to estimate $C_{\ell}$ individually for all the TGSS pointings. This results in estimates of $C_{\ell}$ spanning approximately the  $\ell$ range $100 \le \ell \le 4,000$ in $3893$ different pointing directions. We have removed some pointings due to large system noise or the presence of strong RFIs. The analysis was carried out both before and after source subtraction, with the aim of the analysis being threefold.  
The first aim here is to directly characterize the fluctuations in the sky brightness for different pointing directions on the sky. This provides a direct estimate of the foregrounds for EoR 21-cm observations centred at different directions of the sky.

Source subtraction \citep{ali,ghosh150,beardsley16,gehlot18,kerrigan18} offers a technique for foreground mitigation; this, however, is limited by our ability to accurately calibrate the visibility data and model the sources. The TGSS ADR1 uses a novel method to incorporate Direction-dependent (DD) calibration to all the data sets. It generates a model of the ionosphere using a few strong sources present in that field and corrects the phase due to this ionospheric distortion. It helps to model the extra-galactic point sources with more accuracy. We have removed the discrete point sources above $5\sigma$ $(\sigma$ is below $5 {\rm mJy}$ for a majority of the pointings). The second aim here is to investigate the foreground reduction that is actually achieved through source subtraction in different observing directions of the sky.

Finally, we attempt to use the residual data after source subtraction to quantify the statistical properties of the DGSE, which is expected to be the dominant foreground contribution after source subtraction.  A brief outline of the paper follows. In Section 2, we briefly describe the GMRT data and the method of analysis. In Section 3, we show our measurement of the angular power spectrum before and after point source subtraction and discuss the quantum of drop that occurs in different directions due to the efficacy of point source removal. The comparison with the single-dish observation is presented in Section 4. Finally, we summarize and conclude in Section 5. In a companion paper, we plan to show 
the details of the power-law fitting, the variation of the power-law index and interpretation of the $C_{\ell}$ of the DGSE from residual data.

\section{Methodology}
\label{method}
The TGSS \citep{sirothia14} is the first all-sky continuum survey at a low frequency which is directly relevant for EoR studies. The observing frequency for this survey is $150$ {\rm MHz} with a bandwidth of $16.7$ {\rm MHz}. Although the data were recorded with full polarization, we have used only stokes I for this work. The total survey area is divided into $5336$ individual pointings on an approximate hexagonal grid and the integration time for each pointing is about $15$ min. Here we summarize the methodology from data reduction to power spectrum estimation. We divide the total process into two parts: data analysis and the power spectrum estimation.

The TGSS survey data were analysed by using a fully automated pipeline Source Peeling and Atmospheric Modeling (SPAM) package \citep{intema09,intema14}. This pipeline consists of a pre-processing and a main-pipeline component. The pre-processing part converts the raw data into pre-calibrated visibilities for each pointing. Flagging, gain calibration, bandpass calibrations and also the correction for the system temperature variation are incorporated in this part to improve the quality of the data. Finally, the main-pipeline section converts the pre-calibrated visibility into the final calibrated data set and final stokes-I image for each pointing. Here, both the Direction-independent calibration and Direction-dependent (DD) calibration are applied to the data. The details of the analysis can be found in \citet{intema17}. The background RMS noise is below $5 {\rm mJy}$ for majority of the pointings with an angular resolution $25^{''}\times25^{''}$ (or $25^{''}\times25^{''}$/cos(DEC-19 deg) for pointings south of 19 deg DEC.  The discrete point sources above a $5\sigma$ threshold value have been removed in the final residual data sets.  In this paper, we used both the data before and after point source subtraction to estimate the angular power spectrum.  In our earlier paper (\citealt{samir17a}), we have presented results for the two fields located at the galactic coordinates of $(9^{\circ},+10^{\circ})$ and $(15^{\circ},-11^{\circ})$.  The present work is an extension of our earlier work where we now analyze the entire sky region covered by the TGSS.

In this paper, we have used the Tapered Gridded Estimator (TGE) \citep{samir16b} to estimate the angular power spectrum $C_{\ell}$.  Here, we briefly summarize the salient features of the TGE. The TGE has three main characteristics: (a) it uses the gridded visibility data to reduce the computation. (b) it tapers the sky response from the outer region of the primary beam where it is highly frequency dependent and (c) it subtracts the noise bias to give an unbiased estimate of the true sky signal. We divide the whole ``uv'' plane in a rectangular grid. We have convolved the measured visibilities around each grid point with the Fourier transform of a window function which will effectively taper the sky response. The convolved visibility $\V_{cg}$ at every grid point $g$ can be written as
\begin{equation}
\V_{cg} = \sum_{i}\tilde{w}(\u_g-\u_i) \, \V_i
\label{eq:a2}
\end{equation}
where $\tilde{w}(\u)$ is the Fourier transform of the tapering window function ${\cal W}(\theta)$, $\u_g$ refers to the baseline of different grid points and $\V_i$ is the  visibility measured at
baseline $\u_i$. Here we collapse the visibility measurements in
different frequency channels after scaling each baseline to the
appropriate frequency.  The TGE correlates the convolved visibilities at each grid point to estimate the $C_{\ell}$. As mentioned, it subtracts the self-correlation of the measured visibilities around each grid point which is responsible for the noise bias. The mathematical expression for the TGE is given by (equation 17; \citealt{samir16b}),

\begin{equation}
{\hat E}_g= M_g^{-1} \, \left( \mid \V_{cg} \mid^2 -\sum_i \mid
\tilde{w}(\u_g-\u_i) \mid^2 \mid \V_i \mid^2 \right) \,,
\label{eq:a6}
\end{equation}

where $M_g$ is the normalizing factor which we have calculated by
using simulated visibilities corresponding to an unit angular power
spectrum (details in \citealt{samir16b}). Assuming that the signal is
isotropic, we average the $C_{\ell}$ measurements over an annular region to increase the signal to noise ratio. We use equations (19)
and (25) of \citet{samir16b} to estimate the $C_{\ell}$ and its
variance in bins of equal logarithmic interval in $\ell$. Here, we
divide the whole $\ell$ range in 25 equally spaced logarithmic
bins. The estimator has been already validated using realistic simulations of GMRT $150$ {\rm MHz} observations \citep{samir16b}. In \citet{samir16a} we included point sources in a large region of the sky and showed that the TGE effectively suppresses the point source contribution from outer regions of the primary beam.

\section{Results}
\label{results}
In this section, we show the results for the angular power spectrum measurements before and after subtracting the point sources from the calibrated visibility data. We have removed all fields which are dominated by the system noise or have strong RFI. Finally, out of the $5336$ TGSS fields, we present results for $3893$ fields which we expect to be dominated by the sky signal.

\begin{figure*}
\begin{center}
\includegraphics[width=80mm,angle=0]{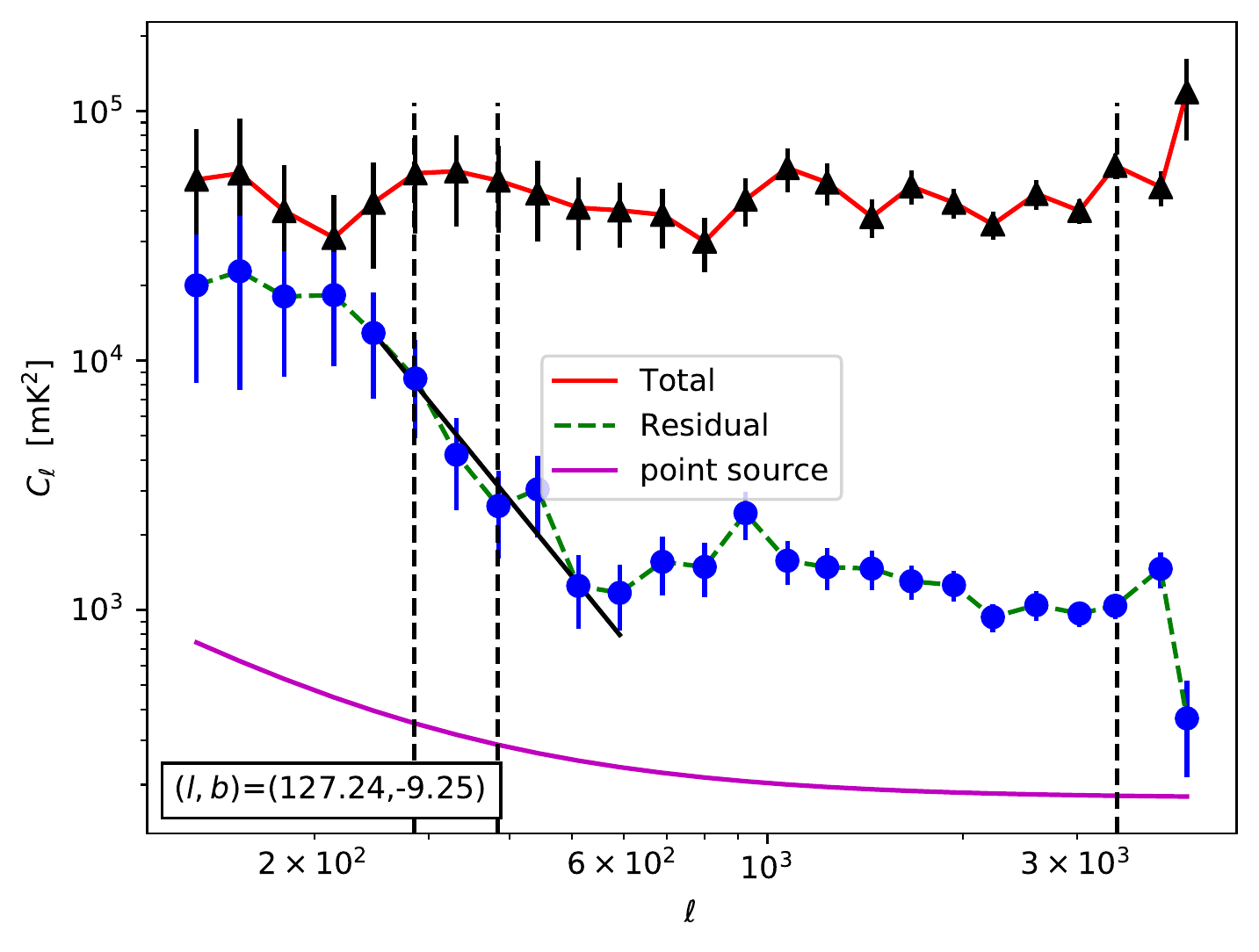}
\put(-60,155){Field1}
\includegraphics[width=80mm,angle=0]{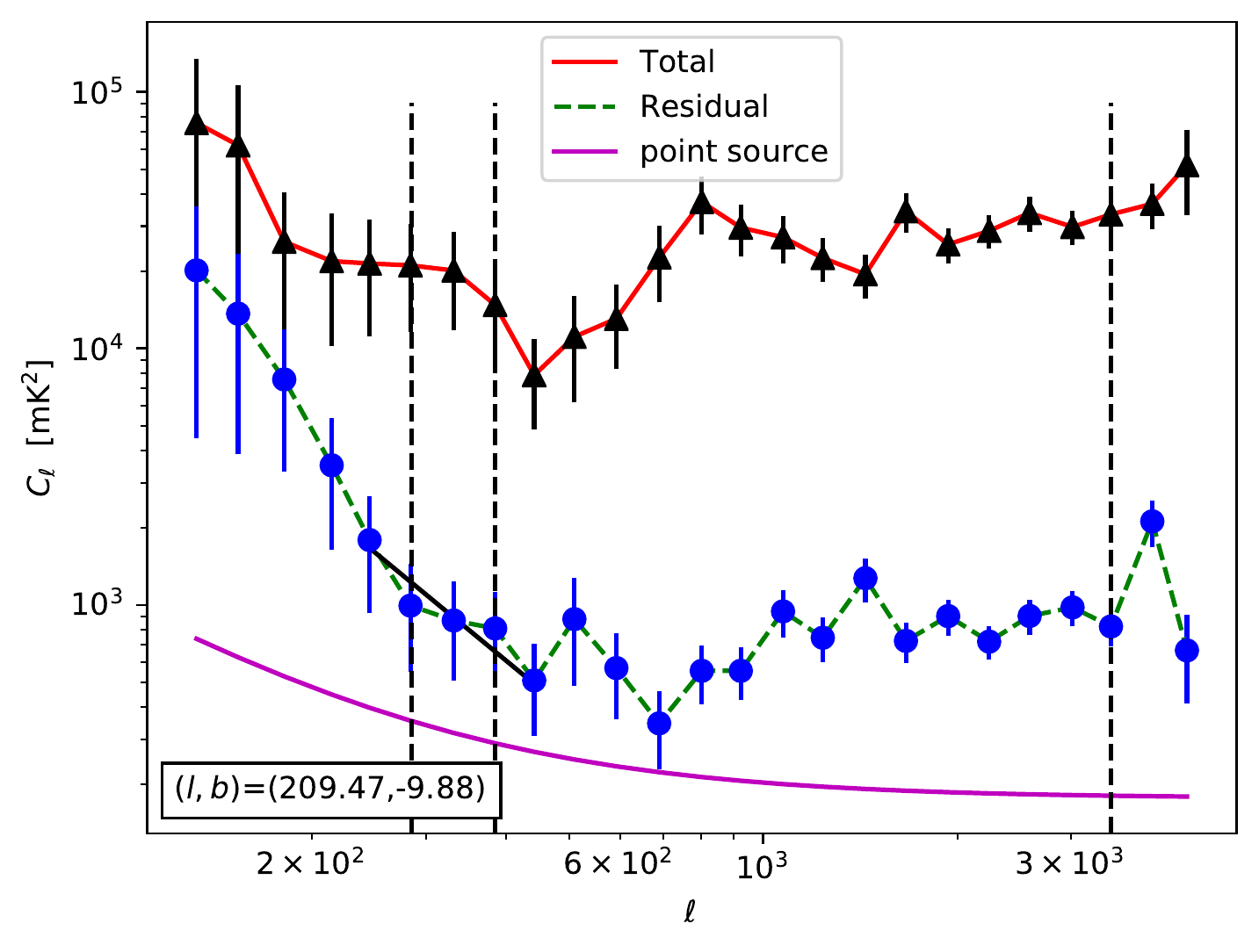}
\put(-60,155){Field2}

\includegraphics[width=80mm,angle=0]{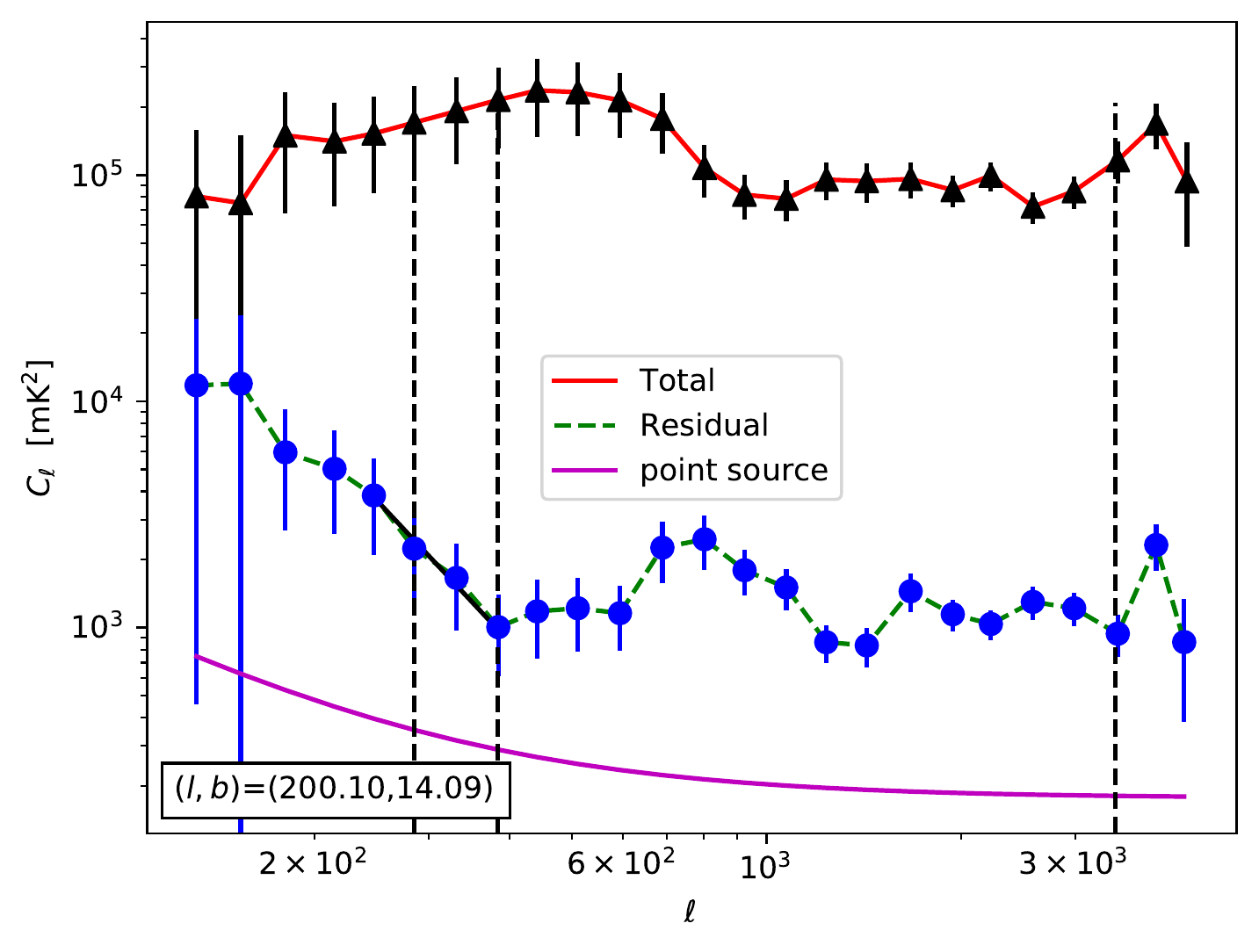}
\put(-60,155){Field3}
\includegraphics[width=80mm,angle=0]{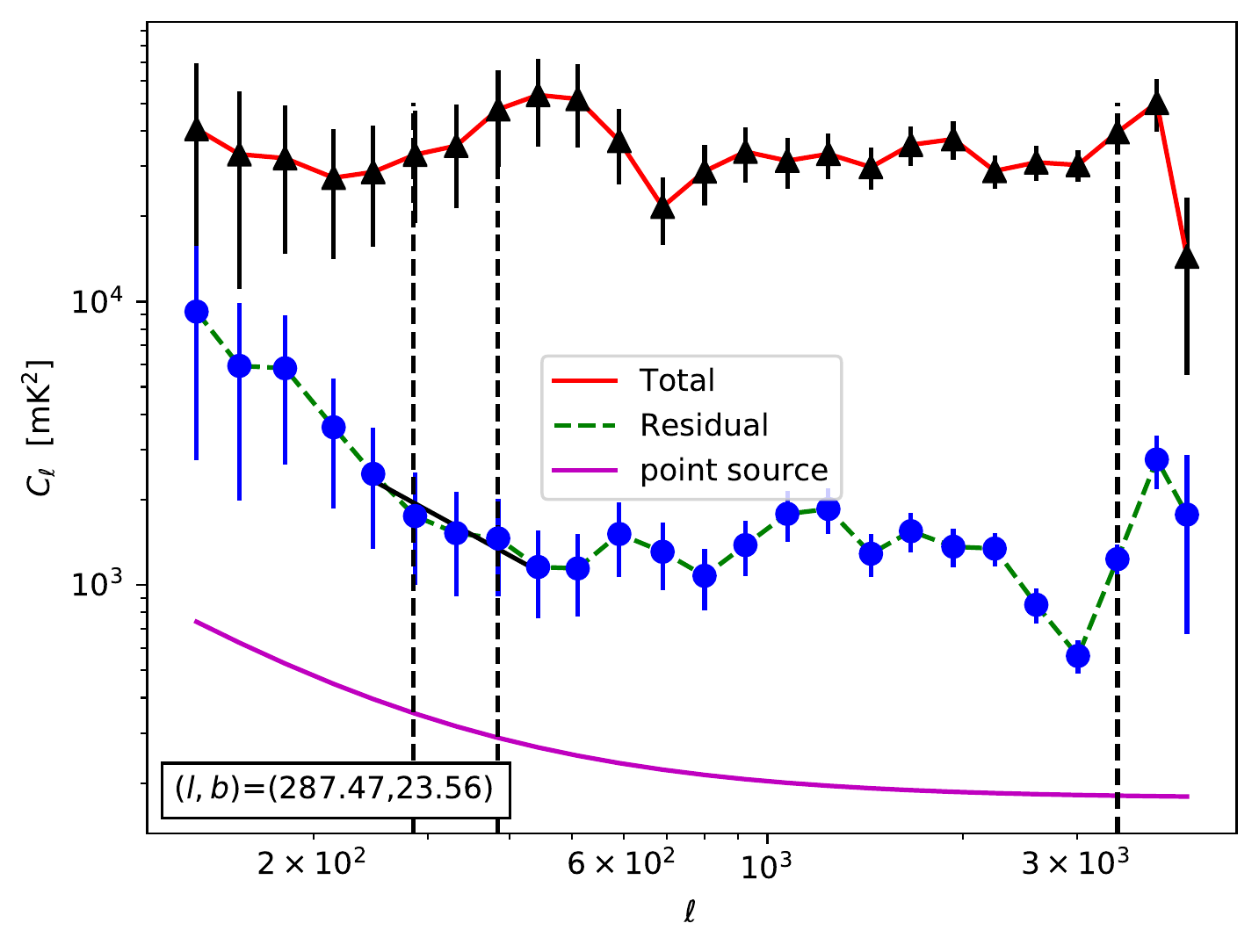}
\put(-60,155){Field4}
\caption{The estimated angular power spectra  $C_{\ell}$ with $1\sigma$ error bars for four representative fields with different galactic coordinates. The upper and lower curves are for before and after point source subtraction respectively. Here, three vertical lines show the values of $\ell$ which we use in Figure \ref{fig:fig2} and Figure \ref{fig:fig2aaa}. The black solid line in each panel shows the best fit power-law for the measured $C_{\ell}$ (details in the text). The magenta line shows the $C_{\ell}$ model prediction due to unsubtracted point sources below $50$ {\rm mJy}.}
\label{fig:fig1}
\end{center}
\end{figure*}

Figure \ref{fig:fig1}  shows the measured $C_{\ell}$ before and after point source subtraction for four representative fields with galactic co-ordinates $({\it l,b})$ = $(127.24,-9.25)$, $(209.47,-9.88)$, $(200.10,14.09)$ and $(287.47,23.56)$ respectively. The results for all of the $3893$ TGSS pointings which have been analyzed here are available online\footnote{\verb|http://www.physics.iisc.ernet.in/~nroy/plot_html|}. The upper curves in these figures show  $C_{\ell}$ with $1\sigma$ error bars before point source subtraction. Here the values of $C_{\ell}$ are in the range $10^4-10^5~{\rm mK^2}$ and the nature of these curves are more or less flat. The sky signal here is predicted \citep{ali} to be dominated by the Poisson fluctuations of the point source distribution, and the nearly flat $C_{\ell}$ is roughly consistent with this prediction.  At low $\ell$, the measured $C_{\ell}$ is affected by the convolution with the tapering window function and the antenna's primary beam pattern. As shown in Figure 2 of \citet{samir17a}, this convolution becomes important at $\ell<240$. Also, the clustering of the point sources starts to become important at the lower $\ell$ values. 

Point sources with flux above a threshold flux $S_{cut}$ were subtracted from the data. Here $S_{cut}=5 \, \sigma$ where $\sigma$ is the rms. noise which varies from pointing to pointing and $ \sigma$ is below $5 \ {\rm mJy}$ for the majority of the pointings \citep{intema17}. The lower curves in Figure \ref{fig:fig1} show the measured $C_{\ell}$ with $1\sigma$ error bars after subtracting the point sources from the data. We see that the values of $C_{\ell}$ at large $\ell$ fall substantially after the point sources are removed. The  $C_{\ell}$ of the residual data shows a nearly flat nature with values $\sim 10^3~{\rm mK^2}$ at $\ell>700$. We believe that this is predominantly the contribution from the Poisson fluctuations of the residual point sources which have fluxes $S < S_{cut}$. The fact that  $C_{\ell}$ at large $\ell$ drops by nearly a factor of $100$ after the point sources are subtracted is a clear indication that the original data is point source dominated. In Figure \ref{fig:fig1}, the magenta line shows the total $C_{\ell}$ prediction due to the clustering and Poisson part of the residual point sources below a threshold flux density of $50$ {\rm mJy}. For this model prediction, we have used source count estimated from \citet{intema17} and the angular correlation function (in the range of $0.1^{\circ}$ to $1^{\circ}$) derived from \citet{dolfi19}. We notice at higher $\ell$, the model $C_{\ell}$ is also dominated by the Poisson fluctuations of residual point sources and the value is an order of magnitude lower than the measured one. We have seen a similar behaviour in our earlier TGGS angular power spectrum analysis \citep{samir17a}. It may be due to that (1.) there are significant residual imaging artefacts around the bright source ($S > S_{cut}$) which were subtracted, and/or (2.) the actual source distribution is in excess of the predictions by TGSS survey at lower flux range. Similar findings have been recently reported at a higher frequency of the 1.28 GHz MeerKAT DEEP2 Image where \citet{Mauch19} found that the model prediction lies significantly below the observed source counts at low flux range. At lower $\ell$ range, we find that the values of the measured $C_{\ell}$ after point source subtraction decrease with increasing $\ell$ and shows a power-law like behaviour (at $\ell<700$). The predicted $C_{\ell}$ due to the clustering point sources at lower $\ell$ values also follows a power law at this range, but the amplitude is, in general, much lower than the residual $C_{\ell}$ (Figure \ref{fig:fig1}). We expect the $C_{\ell}$ measured here to be dominated by the diffuse Galactic synchrotron emission. The TGSS observations can probe the angular scales in the range $0.045^{\circ}$ to $1.2^{\circ}$, however due to convolution of the primary beam and the residual point sources, we are limited in the range $0.3^{\circ}$ to $0.8^{\circ}$ for the DGSE measurements. As discussed in several previous studies \citep{bernardi09,ghosh150,iacobelli13,samir17a}, it is possible to fit $C_{\ell}$ with a power-law $A \times (1000/\ell)^{\beta}$ in this $\ell$ range. We present the results for the power-law fitting in Figure \ref{fig:fig1} for the four representative fields. The black solid line in each panel shows the best fit power law in the $\ell$ range $(\ell_{min},\ell_{max})$. Here we use $\ell_{min}=240$ because convolution becomes important at lower $\ell$ range and $\ell_{max}=600,500,400$ and $450$ for Field 1 to 4 respectively. For $\ell > \ell_{max}$ the residual point sources and other systematic errors dominate and hence we exclude this $\ell$ range in our fitting. The best-fitted values of the parameters are $(A,\beta)=(150\pm56,3.2\pm0.3),(90\pm50,2.1\pm0.4),(53\pm20,3\pm0.3)$ and $(403\pm106,1.3\pm0.2)$ for Field 1 to 4 respectively. The power law fitting we have done over a narrow range, however, that is what we can realistically do with the current data. So, power law power spectrum for the DGSE is an ansatz here, and the power law index ($\beta$) under that assumption is consistent with earlier reported results \citep{samir17a} except for Filed4. The modelling of these four fields are in addition to the two other TGSS fields reported in our earlier paper \citep{samir17a}. The details of the power-law fitting, the variation of the power-law index across different directions in the sky and the interpretation of the residual $C_{\ell}$'s will be presented in a companion paper. For the present purpose, it suffices to note that for the residual data the DGSE dominates the measured $C_{\ell}$ at low $\ell$ ($<700$) whereas the residual point sources dominate at large $\ell$.

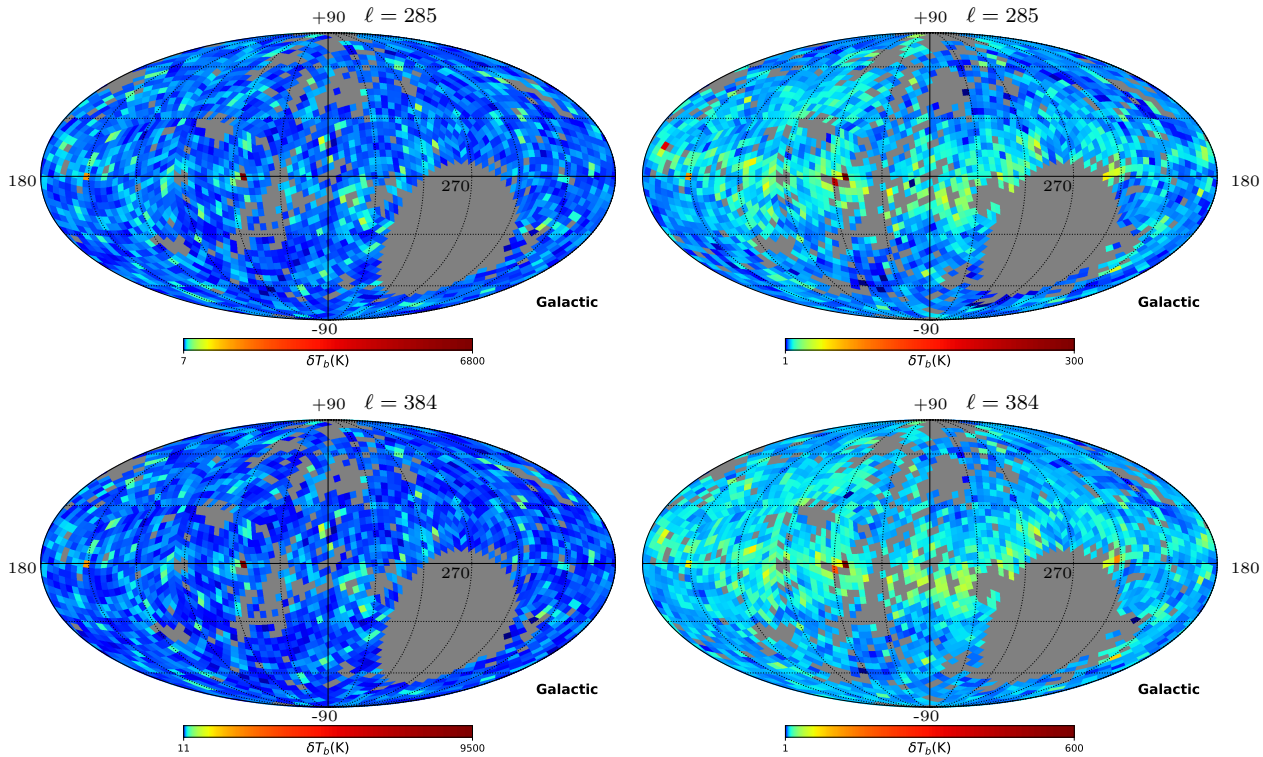
\begin{figure*}
\begin{center}
\includegraphics[width=80mm,angle=0]{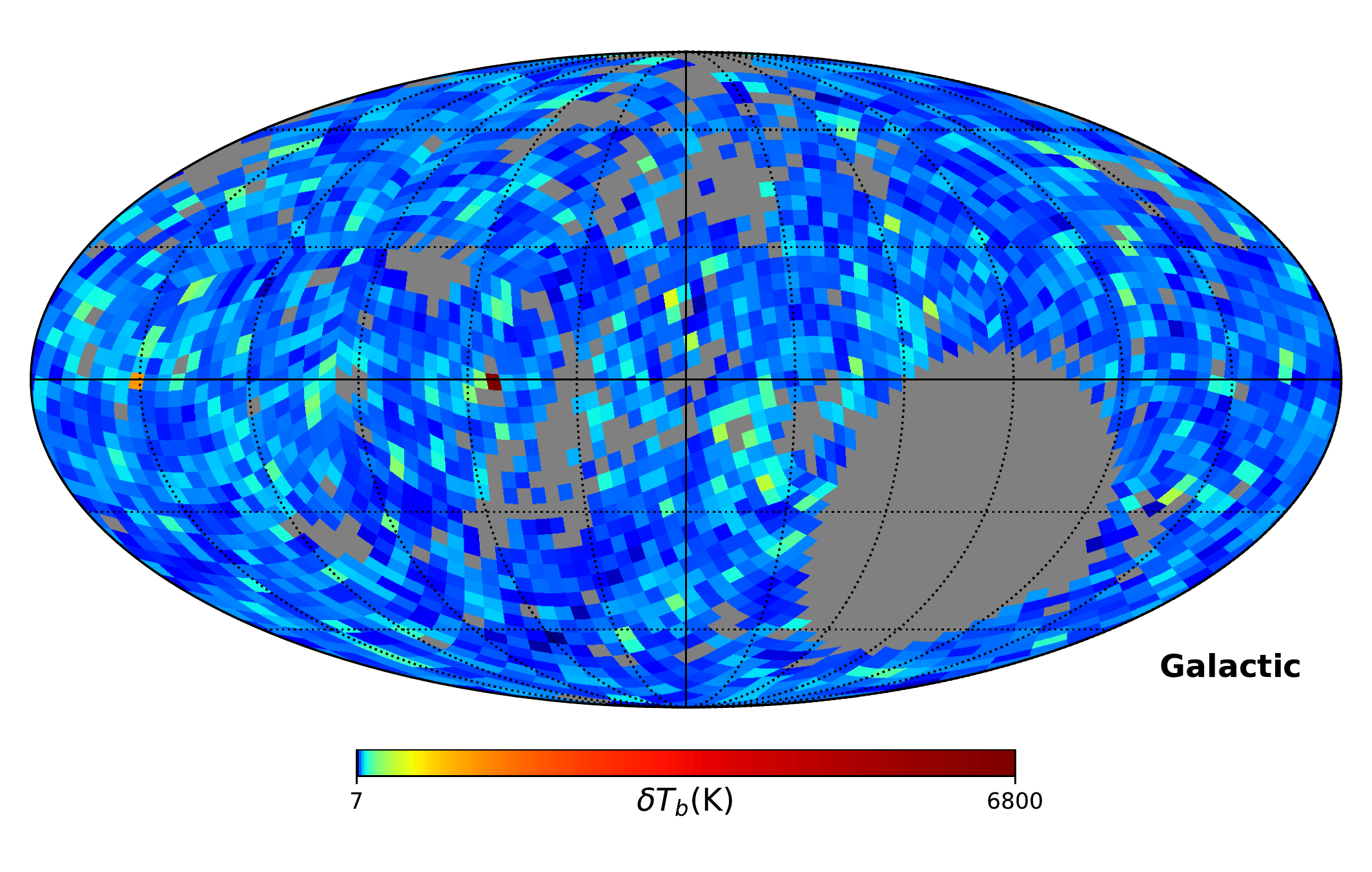}
\put(-235,78){\scriptsize 180}
\put(-120,140){\scriptsize +90}
\put(-120,22){\scriptsize -90}
\put(-71,76){\scriptsize 270}
\put(-100,140){$\ell=285$}
\includegraphics[width=80mm,angle=0]{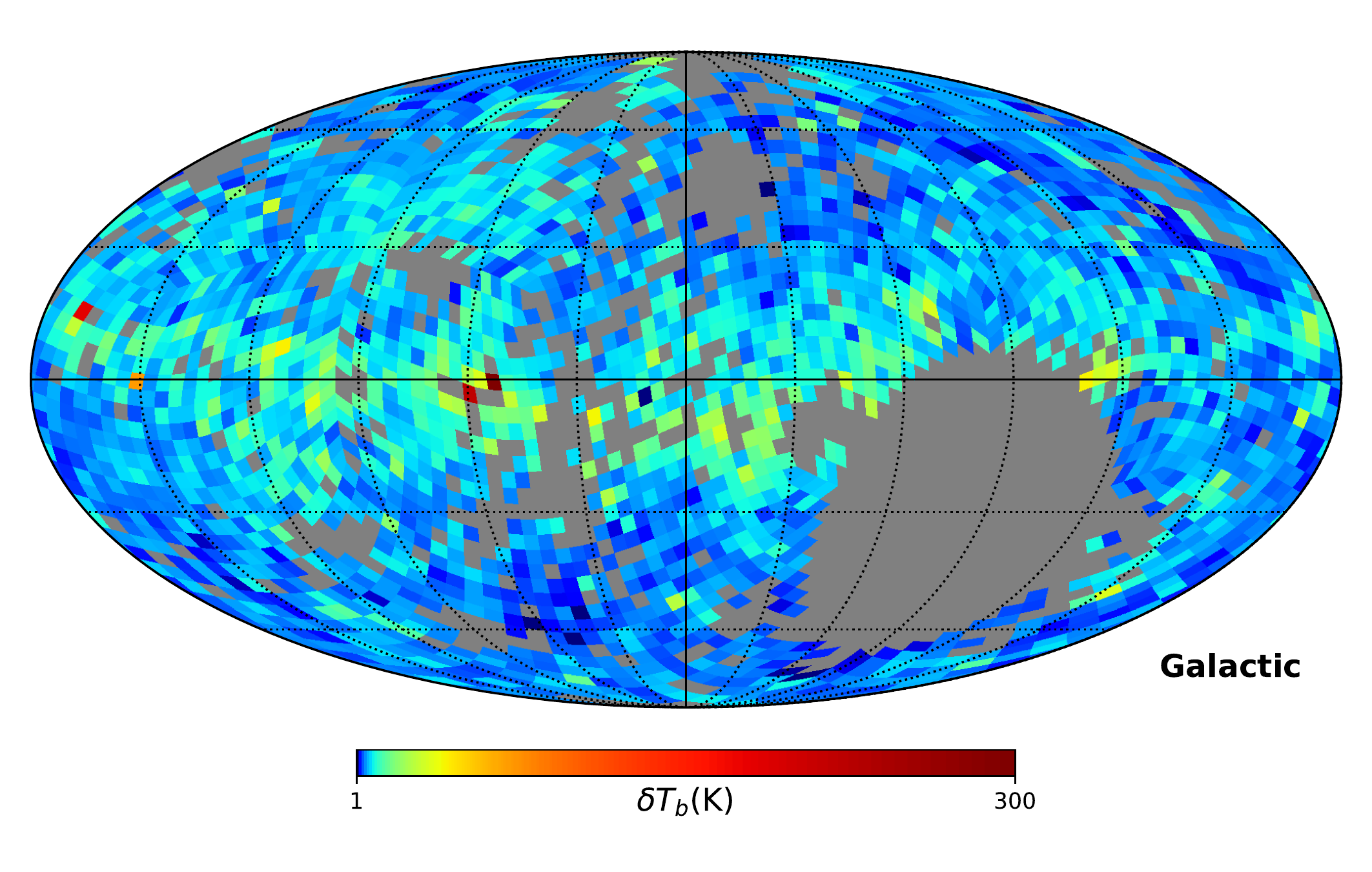}
\put(0,78){\scriptsize 180}
\put(-120,140){\scriptsize +90}
\put(-120,22){\scriptsize -90}
\put(-71,76){\scriptsize 270}
\put(-100,140){$\ell=285$}

\includegraphics[width=80mm,angle=0]{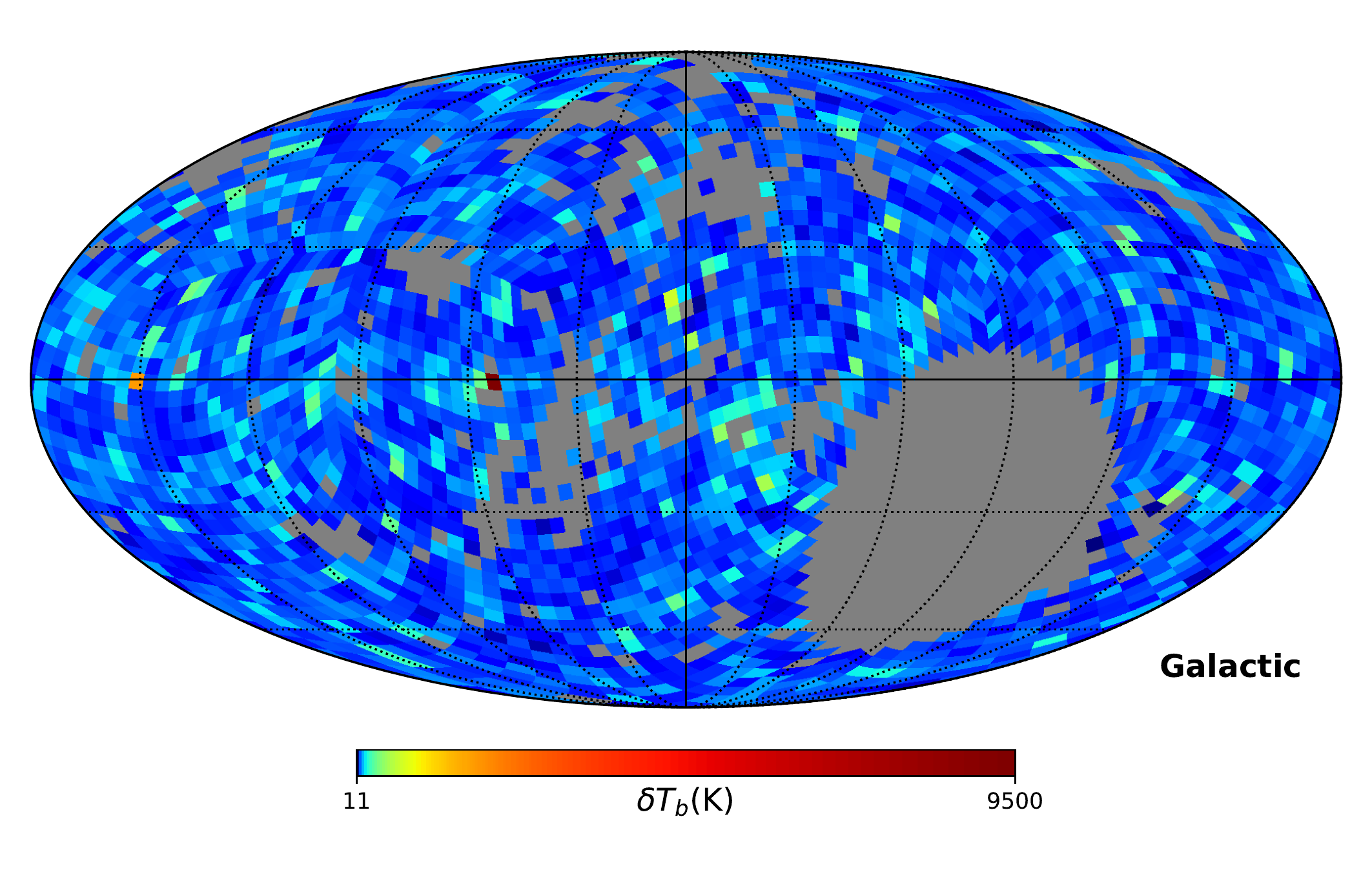}
\put(-235,78){\scriptsize 180}
\put(-120,140){\scriptsize +90}
\put(-120,22){\scriptsize -90}
\put(-71,76){\scriptsize 270}
\put(-100,140){$\ell=384$}
\includegraphics[width=80mm,angle=0]{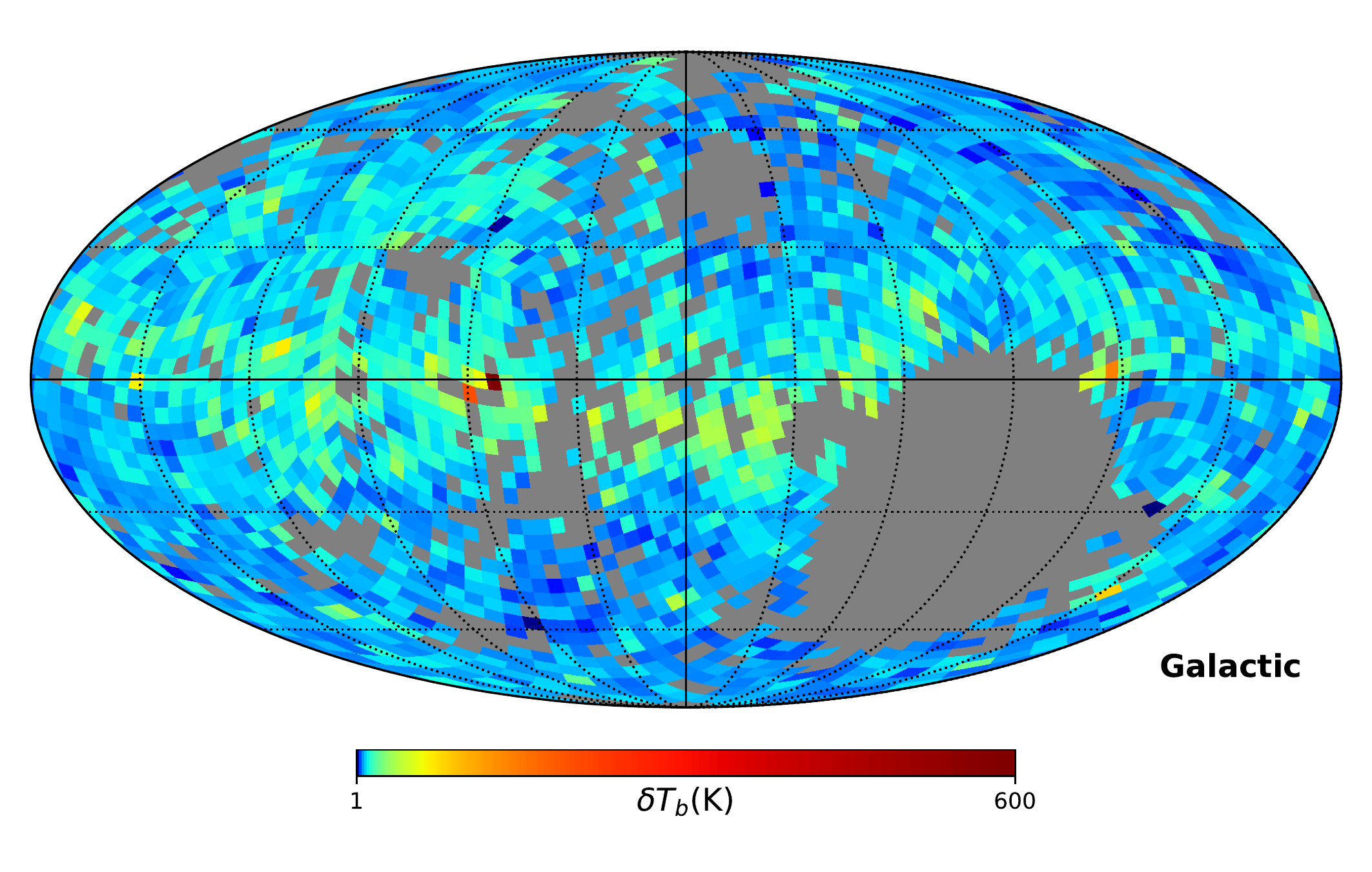}
\put(0,78){\scriptsize 180}
\put(-120,140){\scriptsize +90}
\put(-120,22){\scriptsize -90}
\put(-71,76){\scriptsize 270}
\put(-100,140){$\ell=384$}
\caption{The rms fluctuations of the brightness
  temperature $(\delta T_b)$ all over the sky at different angular scales. Here, the upper and lower panels are for  $\ell=285~(\sim0.63^{\circ})$ and $384~(\sim0.47^{\circ})$
  respectively. The left and right panels show the values of $\delta T_b$ before and after subtracting the point sources from the data.}
\label{fig:fig2}
\end{center}
\end{figure*}

We next consider the rms fluctuations of the brightness temperature $\delta T_b=\sqrt{\ell(\ell+1)C_{\ell}/2\pi}$ at different $\ell$ values each of which corresponds to a different angular scales. Figure \ref{fig:fig2} shows how $\delta T_b$ varies across different pointing directions in the sky. Here, we take the mean $\delta T_b$ for all the TGSS pointings which fall into a particular HEALPix\footnote{\url{http://healpix.sourceforge.net}}\citep{gorski05} pixel. The upper and lower panels correspond to $\ell=285~(\sim0.63^{\circ})$ and $384~(\sim0.47^{\circ})$ respectively,  whereas the left and right panels respectively correspond to before and after point source subtraction. For both the multipoles shown here, we expect the signal to be dominated by the DGSE after point source subtraction. The grey circular regions in the lower right part of these images have no data points as they correspond to the declination range $({\rm Dec} < -53^{\circ})$  which is not covered by the TGSS. We also note a few grey pixels distributed throughout the images were discarded as these correspond to pointings which have not been included in our analysis either due to large system noise or RFIs. In the left panels of Figure \ref{fig:fig2}, we notice that the distribution of $\delta T_b$ is almost isotropic with values in the range of a few hundred Kelvin. The $\delta T_b$ here is mainly due to the Poisson fluctuations of the extra-galactic point sources. The sources, being cosmological in origin, are expected to have an isotropic distribution on the sky. In contrast, considering the right panels which show the values of $\delta T_b$ for the residual data where all the discrete point sources have been removed we see that the  $\delta T_b$ values are somewhat larger near the  Galactic plane and they fall off away from the Galactic plane. The values of $\delta T_b$ vary in the range of a few tens of Kelvin. We believe that the residual $C_{\ell}$ is most likely to be dominated by the Galactic synchrotron emission. However, in the Galactic plane, there will also be an additional contribution from the residual thermal emission from HII regions, and also the residual non-thermal emission from supernova remnants.

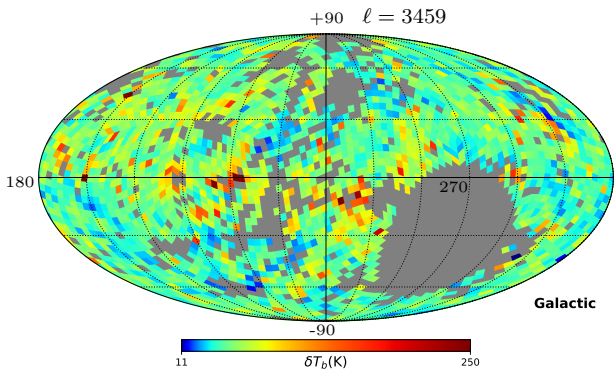
\begin{figure}
\begin{center}
\includegraphics[width=80mm,angle=0]{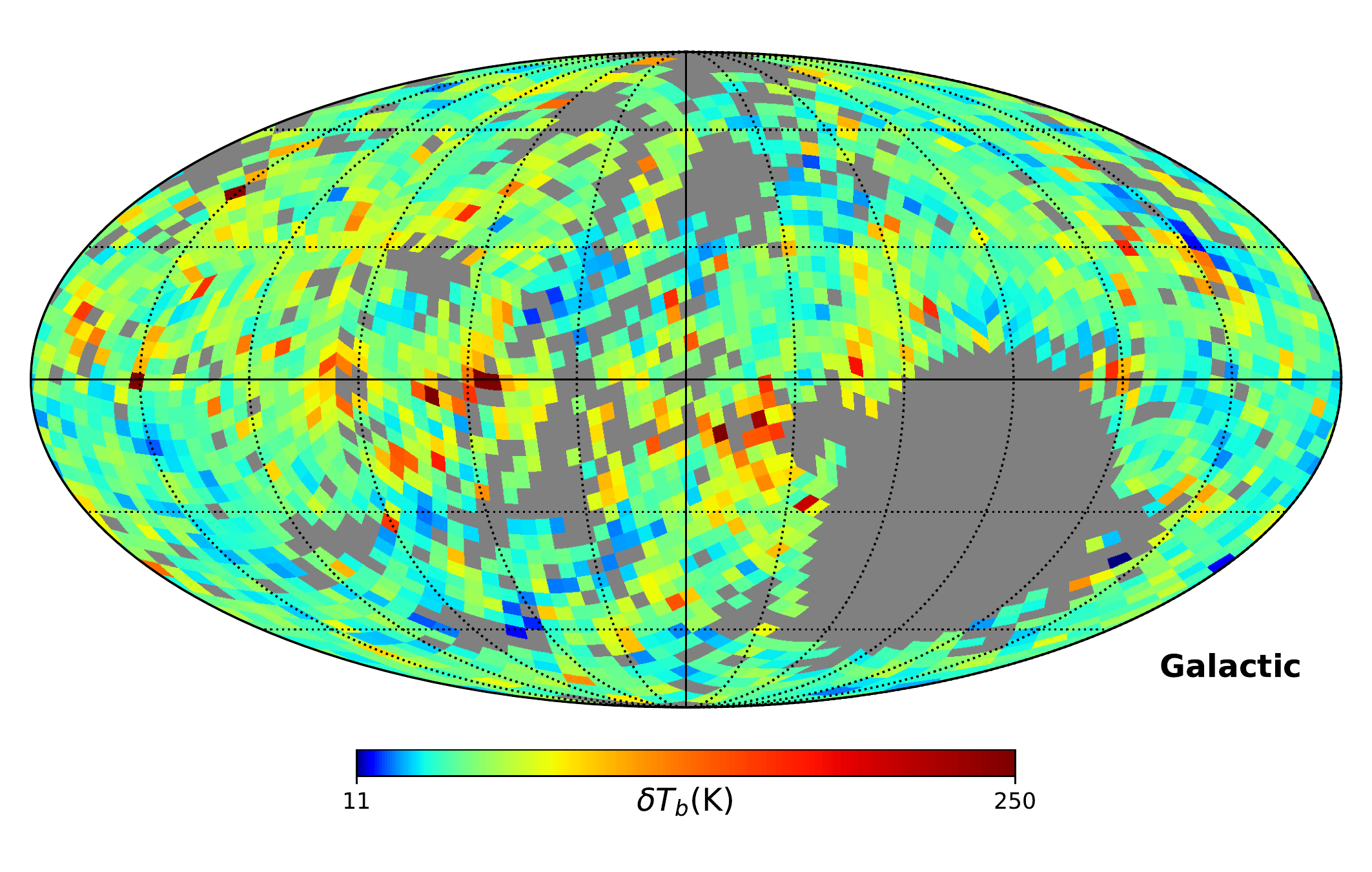}
\put(-235,78){\scriptsize 180}
\put(-120,140){\scriptsize +90}
\put(-120,22){\scriptsize -90}
\put(-71,76){\scriptsize 270}
\put(-100,140){$\ell=3459$}
\caption{The distribution of the $\delta T_b$ at
  small angular scale $\ell=3459~(\sim0.052^{\circ})$. The values $\delta T_b$ are almost isotropic varying mostly within a range of $50-100~{\rm K}$. Few bright pixels in the map are due to the presence of strong bright A-team sources which were not properly modelled and removed.}
\label{fig:fig2aaa}
\end{center}
\end{figure}

The measured $C_{\ell}$ for the DGSE falls as a power law $(C_{\ell}\propto\ell^{-\beta})$ \citep{bernardi09,ghosh150}. In our ealier study with TGSS, we observed the same power law nature of the $C_{\ell}$ as a function of $\ell$ \citep{samir17a}. In \citet{samir17a}, we have also found that the amplitude of the residual $C_{\ell}$ becomes almost constant at $\ell>\ell_{max}(\sim550)$. We believe this angular multipole range $(\ell>\ell_{max})$ is mainly dominated by the the Poisson fluctuations of the residual point sources with flux values below $S_{cut}$. In the four panels of Figure \ref{fig:fig1}, the nature of the all residual $C_{\ell}$ (green dotted curves) are almost similar as in \citet{samir17a}. We can use the residual $C_{\ell}$ at large $\ell$ to set an upper limit of the DGSE at smaller angular scales. In Figure \ref{fig:fig2aaa}, we show the variation of $\delta T_b$ at $\ell=3459~(\sim3^{'})$ after subtracting the point sources. We find the residual map is almost isotropic at $\ell=3459$ and the derived brightness temperature varies in the range $10-100~{\rm K}$. In comparison, the right panels of Figure \ref{fig:fig2} shows the values $\delta T_b$ which are somewhat larger at Galactic plane. This is due to the DGSE which dominates at those $\ell$ values $(\ell=285$ and $383)$ shown in that figure. Also, there are few bright pixels in  Figure \ref{fig:fig2aaa} which might be due to the deconvolution error associated with bright A-team sources in the sky $(>300~{\rm Jy})$ (such as Cas A (${\it l}=111.734, b=-02.129$), Cygnus A (${\it l}=76.1898, b=+05.755$), Hydra A (${\it l}=242.925, b=+25.092$) etc. \citep{intema17}). As mentioned earlier, the residual contribution from the HII regions and supernova remnants may also contribute in the Galactic plane.

%We assume that the measured $C_{\ell}$'s at a particular galactic latitude range are an independent realization of an underlying statistical distribution. Here, each $C_{\ell}$ is an average of square of visibilities over a bin. So, if we assume the measured visibility as a Gaussian random variable, the distribution of the $C_{\ell}$ will follow a $\chi^2$ distribution with a degrees of freedom equal to the number of measurements in that bin. In reality, this number for a particular bin is different at different pointings, so it is really difficult to comment about the exact the statistical distribution of the measured $C_{\ell}$ for a specific latitude range. Here, we quantify the statistical distribution of $C_{\ell}$ through the histogram of the $C_{\ell}$ values.

\begin{figure*}
\begin{center}
\includegraphics[width=42mm,angle=0]{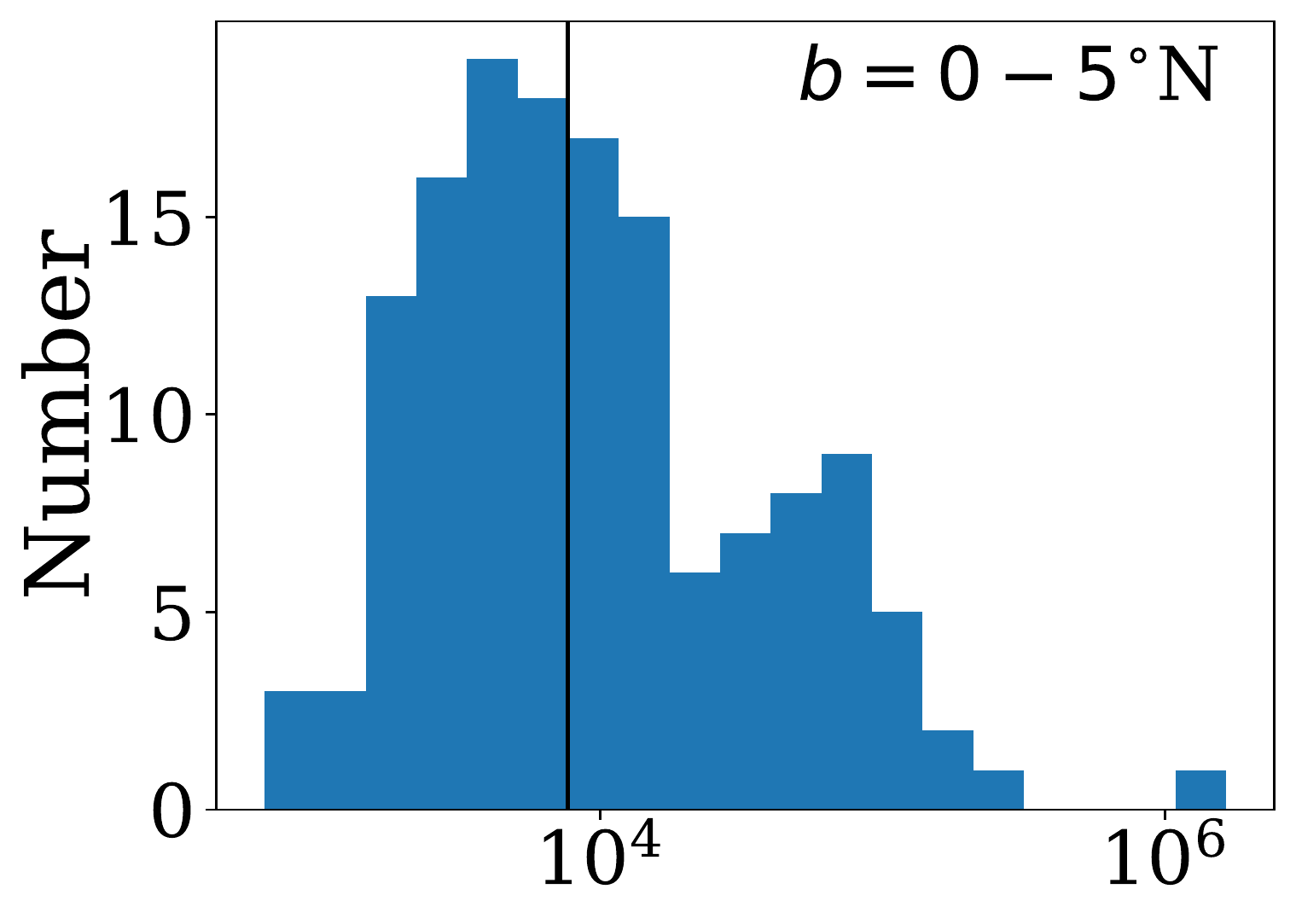}
\includegraphics[width=40mm,angle=0]{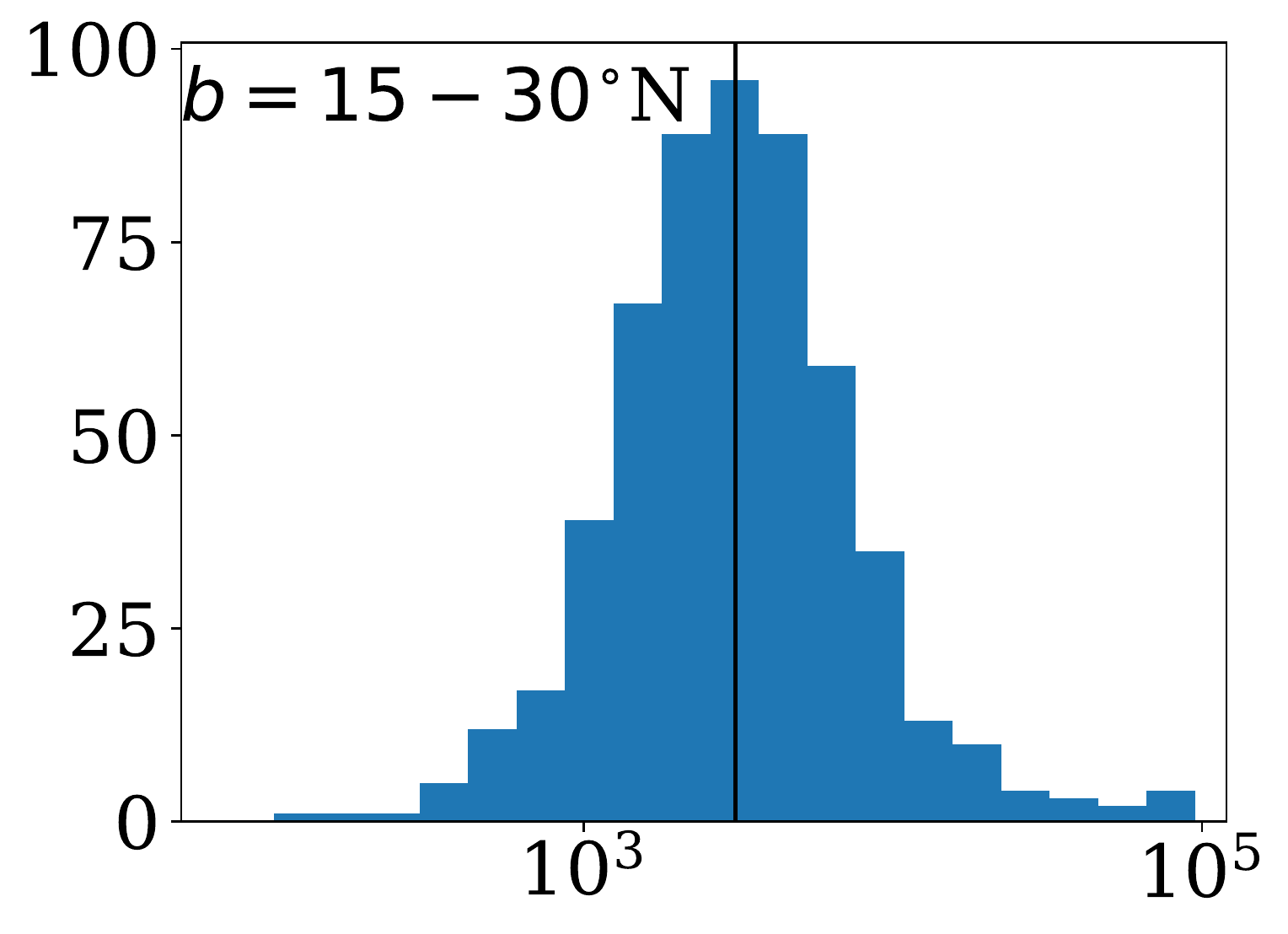}
\includegraphics[width=40mm,angle=0]{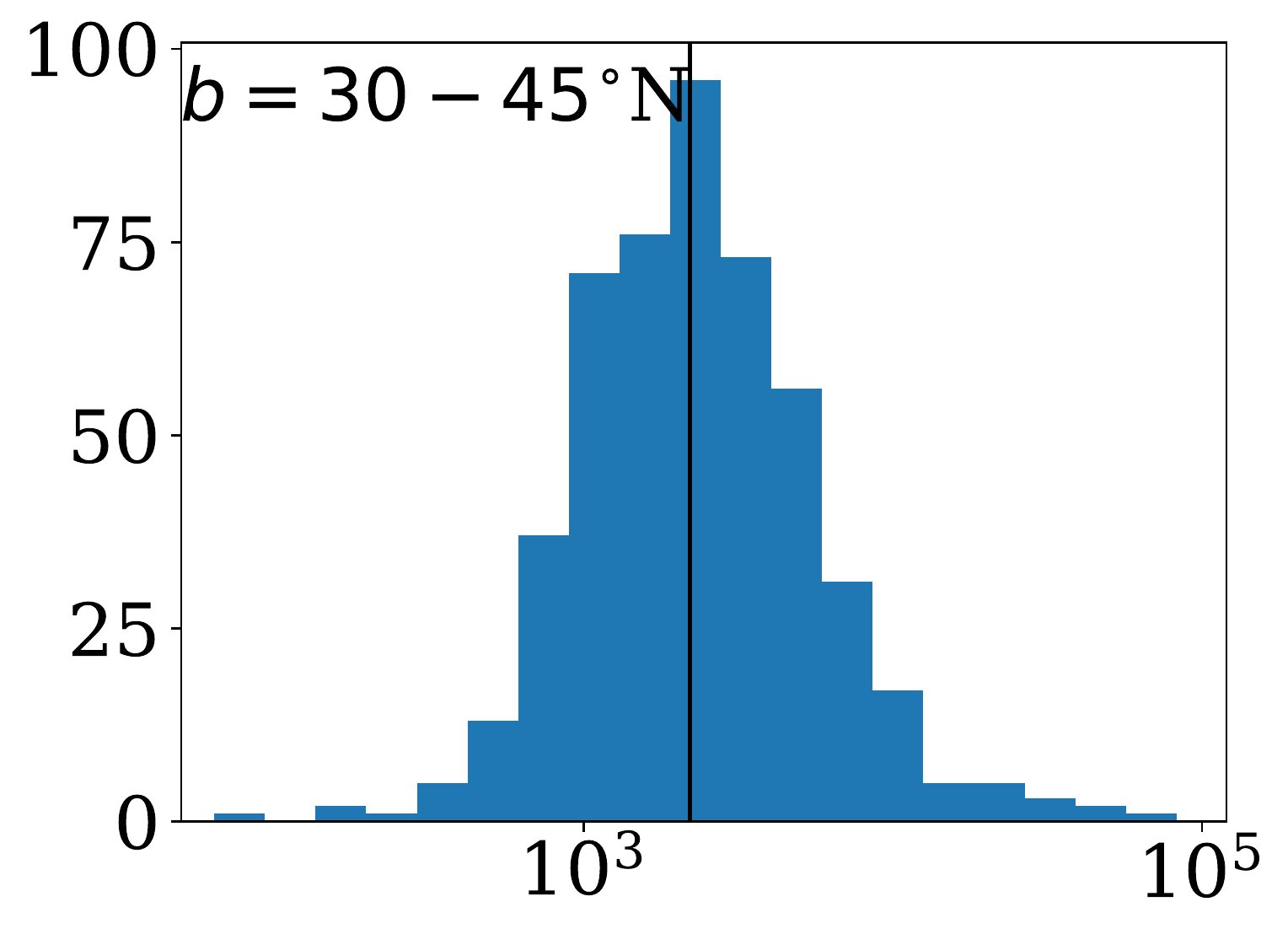}
\includegraphics[width=40mm,angle=0]{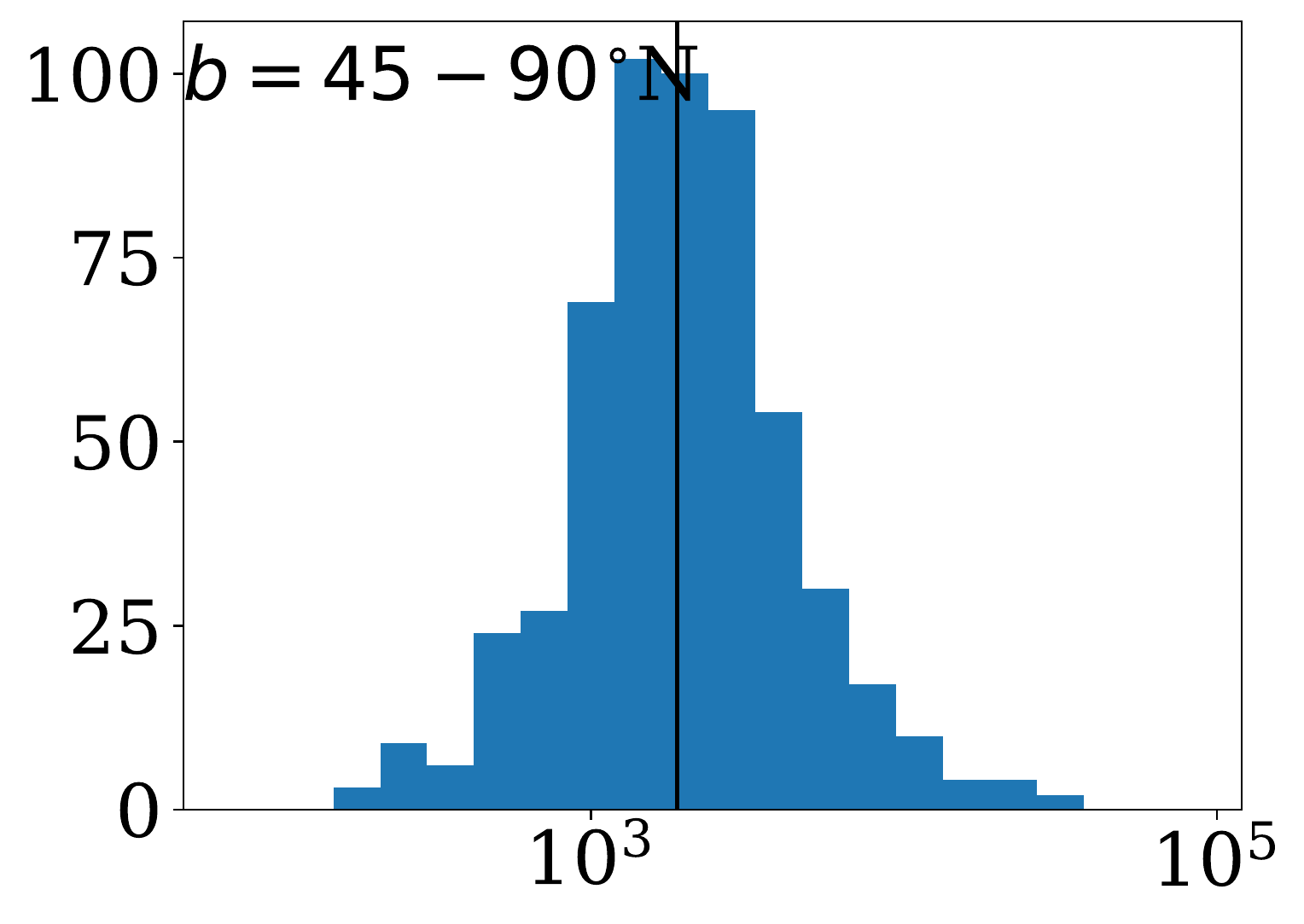}
\includegraphics[width=44mm,angle=0]{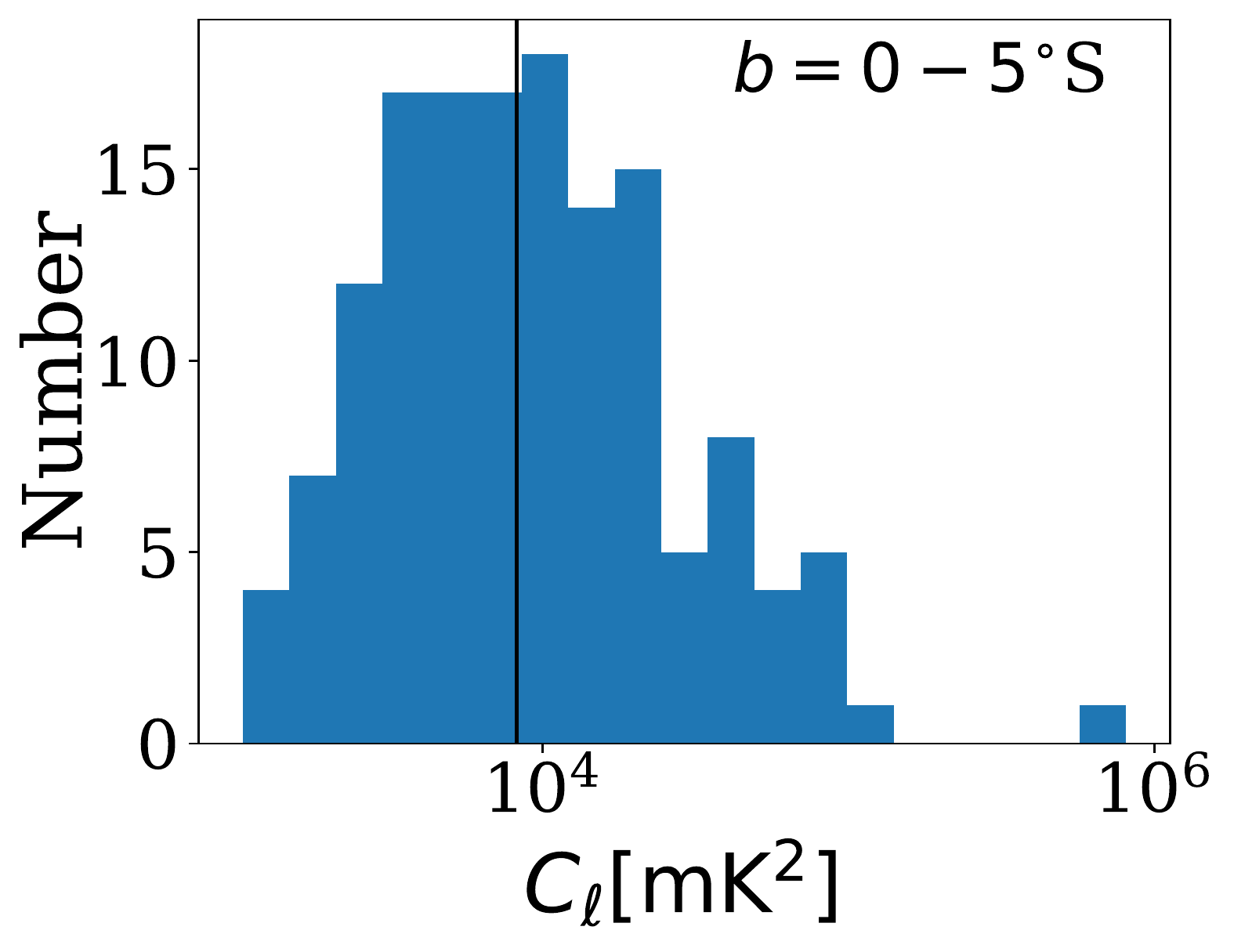}
\includegraphics[width=40mm,angle=0]{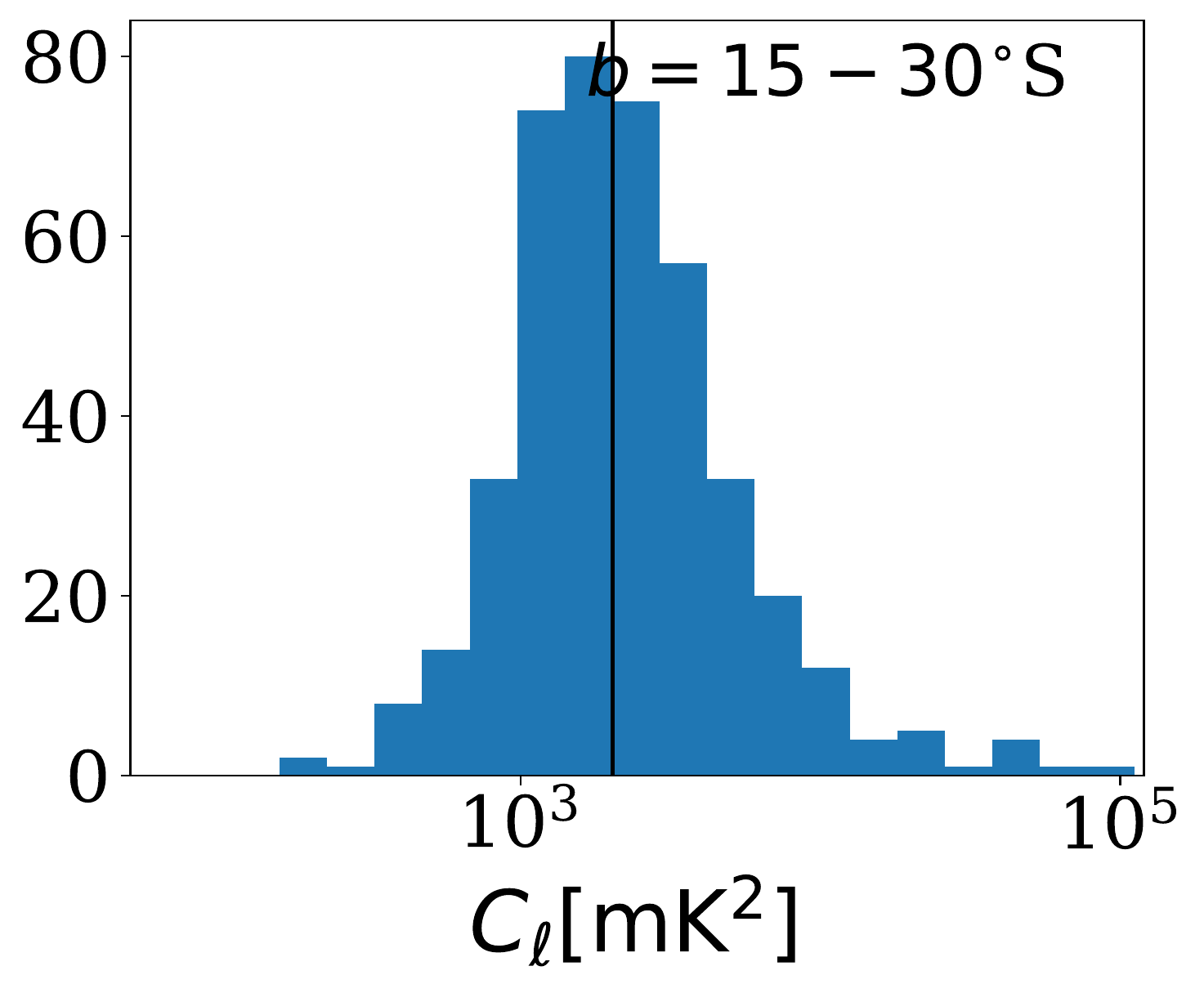}
\includegraphics[width=40mm,angle=0]{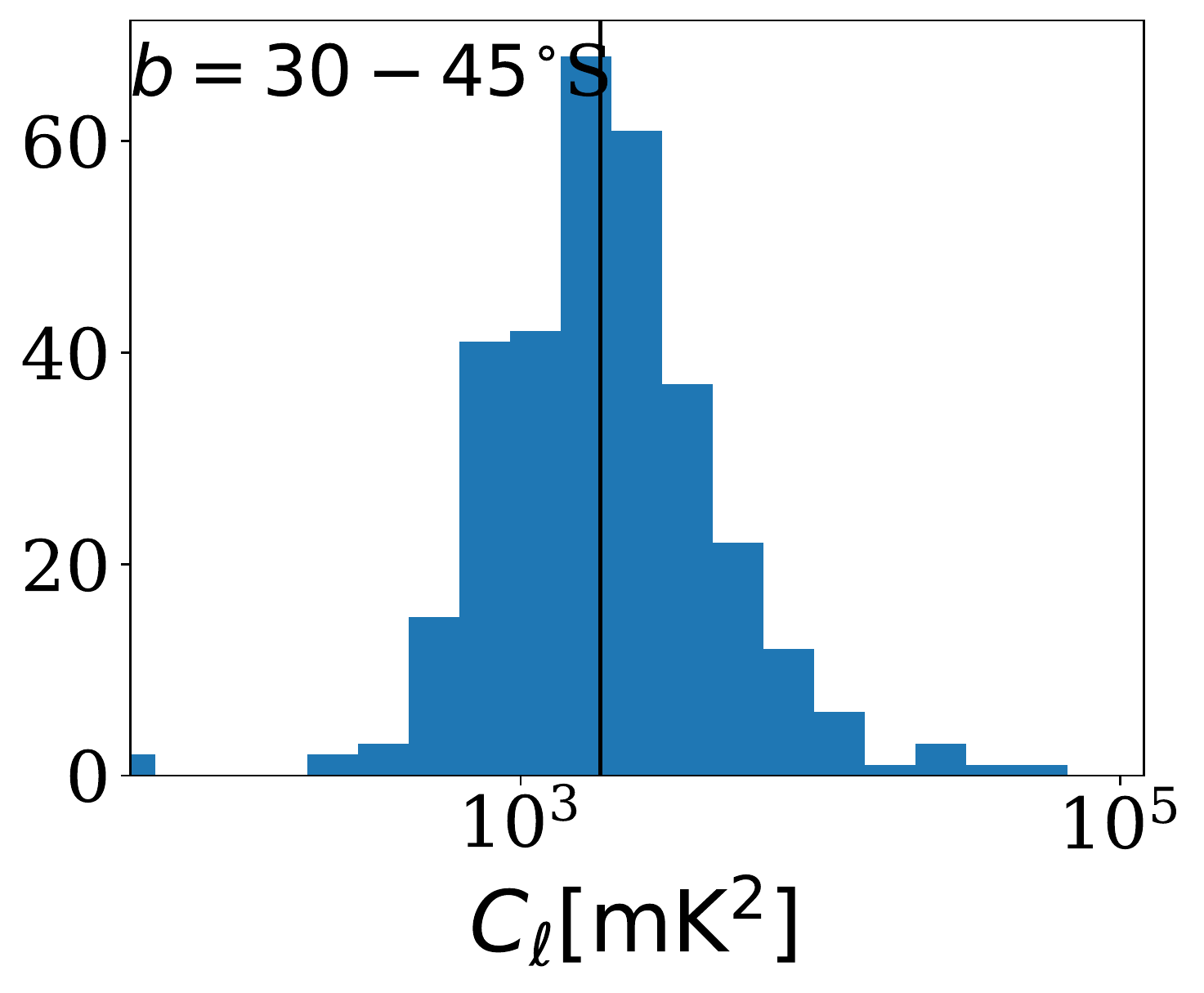}
\includegraphics[width=40mm,angle=0]{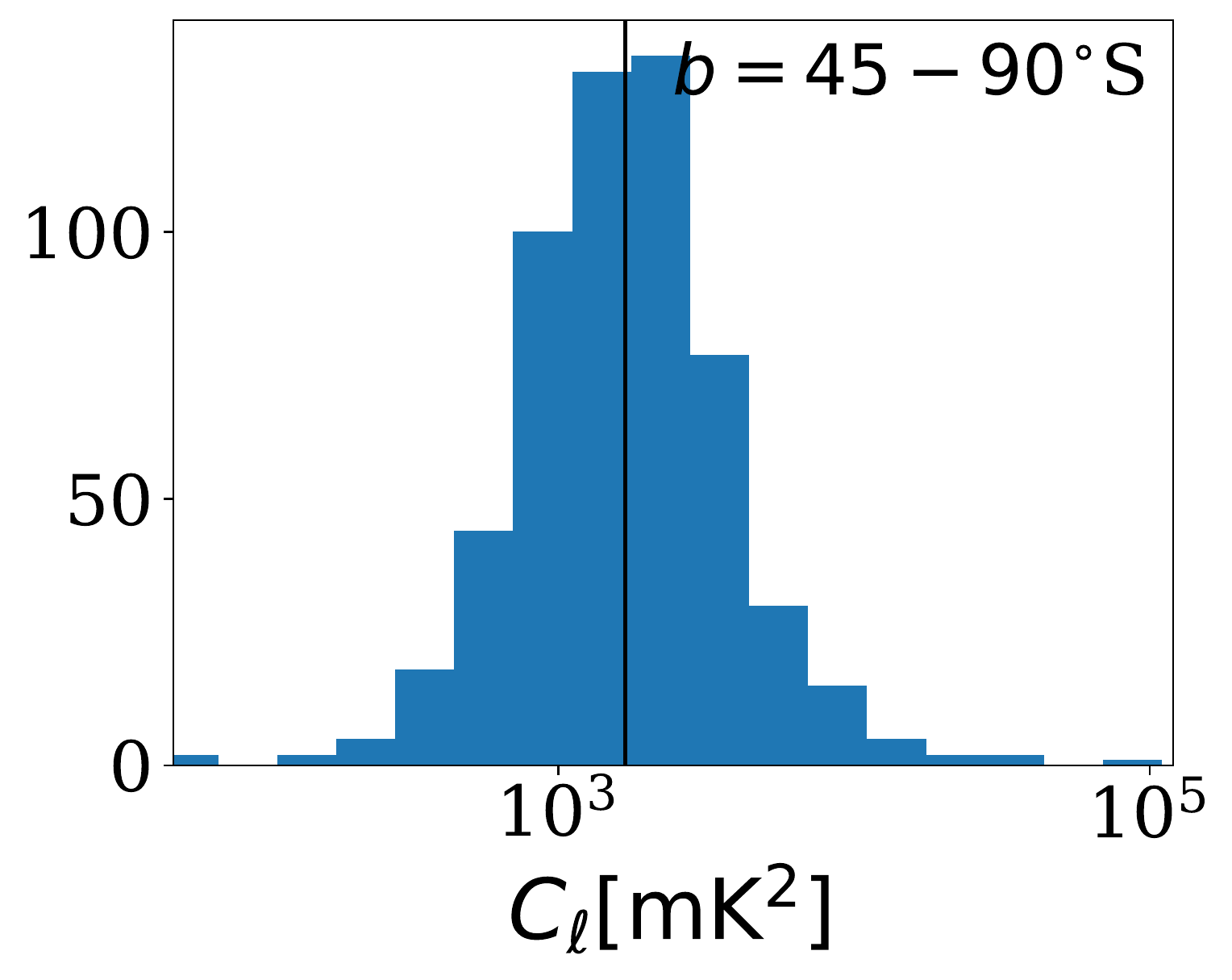}
\caption{Here, we display the histogram of the $C_{\ell}$ values at
  different galactic latitude ranges. The upper and lower panels are for the northern and southern hemisphere respectively. Here we show the distribution for a fixed $\ell=384$. The median values of the   $C_{\ell}$'s are shown by the black vertical lines in each plane.}
\label{fig:fig2aa}
\end{center}
\end{figure*}

 We assume that the measured $C_{\ell}$'s at a particular galactic latitude range are an independent realization of an underlying statistical distribution. Here, we quantify the statistical distribution of $C_{\ell}$ through the histogram of the $C_{\ell}$ values. In Figure \ref{fig:fig2aa}, we plot the histogram of the measured $C_{\ell}$ at different Galactic latitude range. We considered only $\ell=384$, which is mostly dominated by the DGSE. Here, we showed the results for four different latitude range $0-5^{\circ}$, $15-30^{\circ}$, $30-45^{\circ}$ and $45-90^{\circ}$ across the north and south Galactic plane. The median values of the $C_{\ell}$ for each galactic latitude range are shown by the black vertical lines in each panel. We find that the histogram is mostly peaked around the median values, and we do not find the signature of ``long-tailed'' distribution across all the latitude values. We observed that the distributions of the $C_{\ell}$'s are almost similar for all the latitude ranges except at $0-5^{\circ} S/N$. Here, we find the distribution is slightly bi-modal. This may be an artefact of the complex extended sources present in the Galactic plane which were not properly removed from the data. We found the median values are around $\sim7.5\times10^3~{\rm mK^2}$ in the Galactic plane (${\it b}=0-5^{\circ}$) and it falls to $\sim 3\times 10^3~{\rm mK^2}$ as we move beyond the Galactic plane (${\it b}=15-30^{\circ}$). For latitude range ${\it b}>30^{\circ}$, the DGSE becomes much weaker as compared to that on the Galactic plane, and the corresponding median $C_{\ell}$ values are mostly similar across a wide range of latitudes. As noted earlier, here we are mostly dominated by residual point sources and measured $C_{\ell}$'s corresponds to an upper limit of DGSE. Subsequently, we investigate how the median $C_{\ell}$ changes with $\ell$ for different Galactic latitude ranges in north and south hemisphere.

\begin{figure*}
\begin{center}
\includegraphics[width=80mm,angle=0]{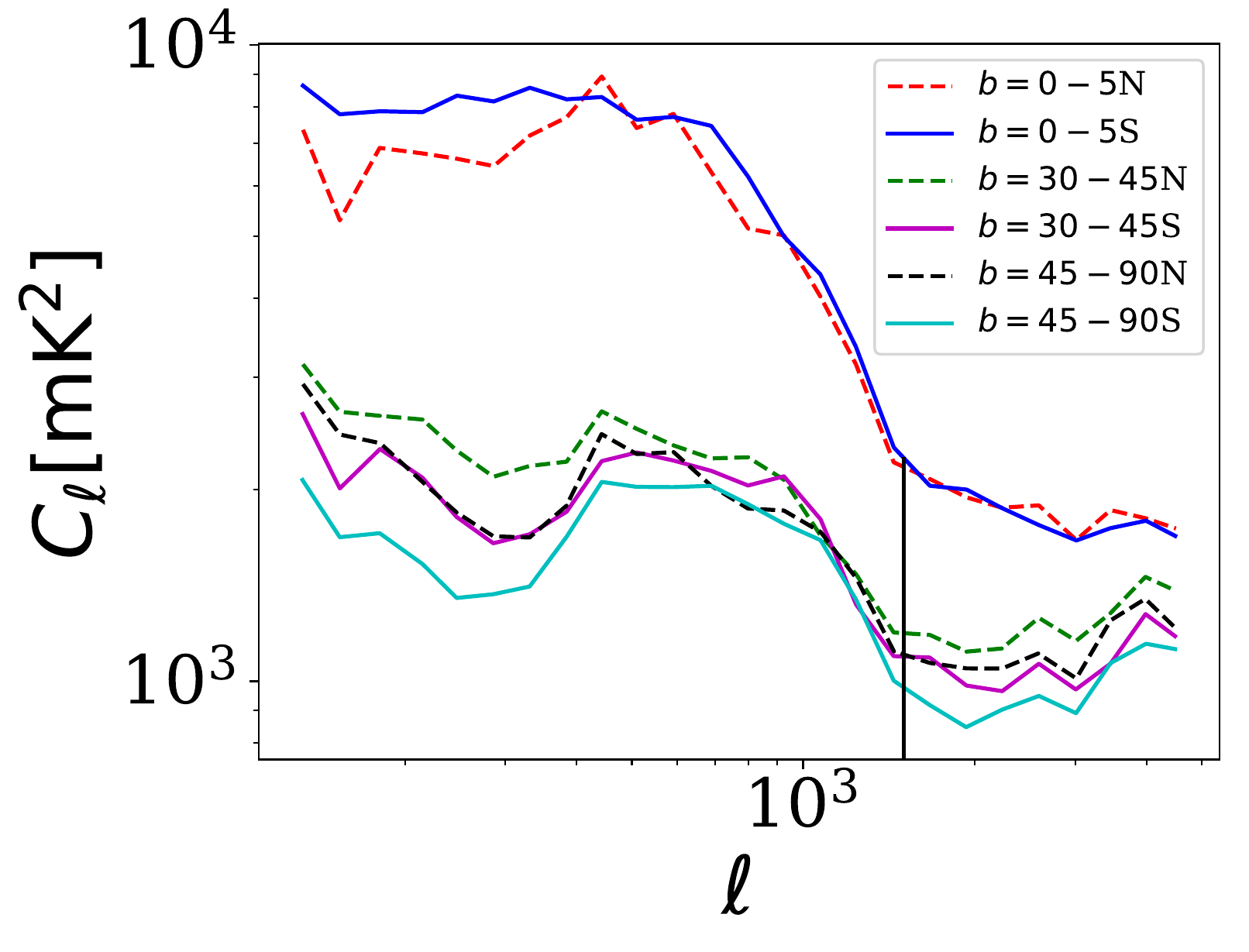}
\includegraphics[width=80mm,angle=0]{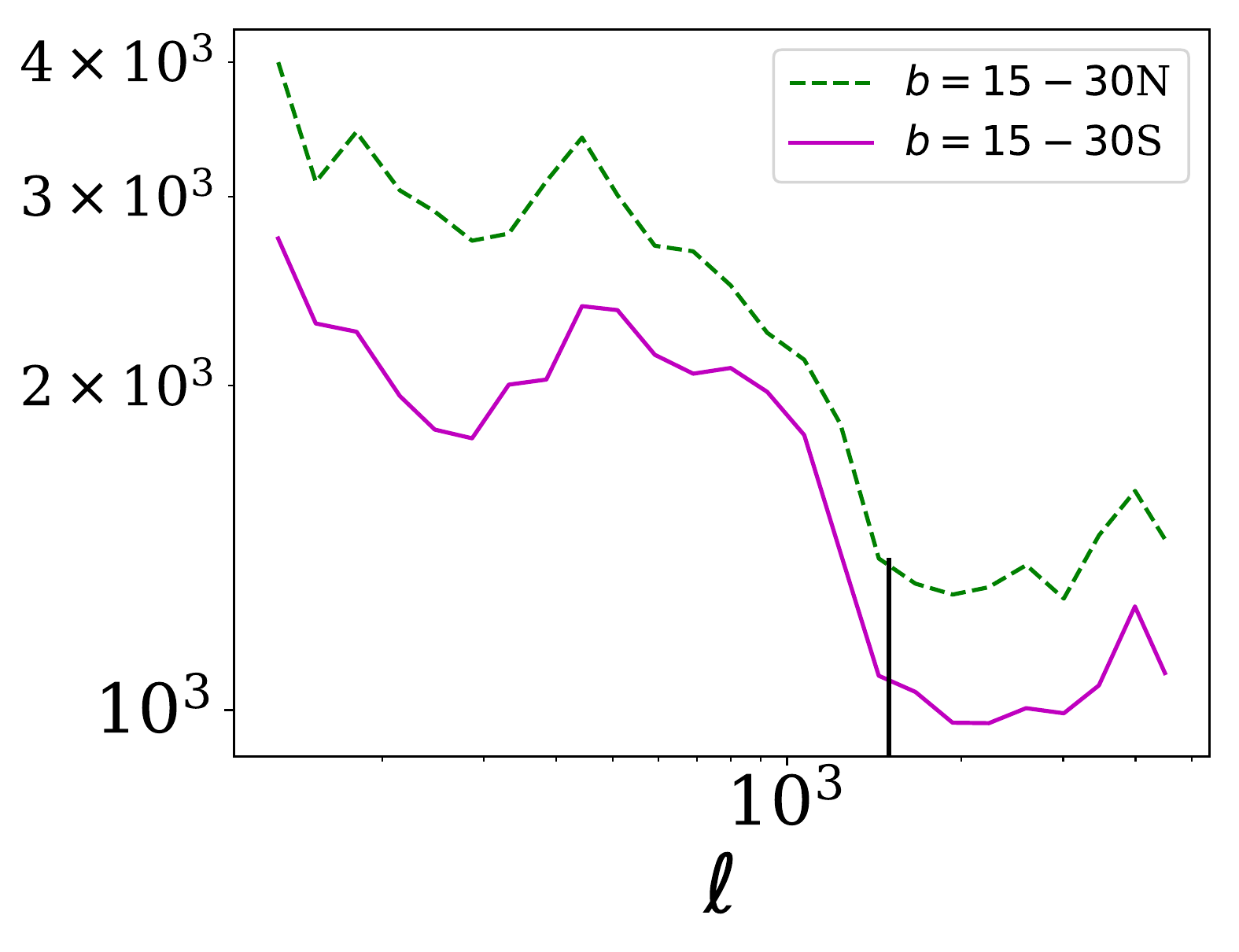}
\caption{This figure compares the median values of the $C_{\ell}$ measurements for the Northern (dashed lines) and Southern hemispheres (solid lines). In left  panel, we show the median $C_{\ell}$ as a function of $\ell$ for latitude range ${\it b}=0-5^{\circ},30-45^{\circ}$ and $45-90^{\circ}$. Here, the measurements are almost symmetric for these two hemispheres. The right panel shows the comparison in latitude range ${\it b}=15-30^{\circ}$. In this case, the measured $C_{\ell}$ are slightly asymmetric, and the overall amplitude is slightly higher for the northern hemisphere. This may be due to the complex structure of disk contributing asymmetrically, or variation of structures due to disk halo interaction in the two hemispheres leading to asymmetric structures in density and magnetic fields.}
\label{fig:fig2a}
\end{center}
\end{figure*}

To quantify the possible signature of North and South asymmetry, we considered the measurements of the $C_{\ell}$ in the Northern and Southern hemisphere. We divide the whole latitude range into different parts and compare the median of the $C_{\ell}$ values. The left and right panels of Figure \ref{fig:fig2a} show the variation of the median $C_{\ell}$ as a function of $\ell$ for different latitude ranges. The blue solid line and the red dashed lines in the left panel show the median $C_{\ell}$ for galactic latitude range $0-5^{\circ}$ for northern and southern hemispheres respectively. We find that the median values are almost symmetric for both hemispheres. The results are also very similar for other latitude ranges ($30-45^{\circ}$ and $45-90^{\circ}$). However, we see some asymmetry in the northern and southern hemisphere in the latitude range $15-30^{\circ}$ (right panel of Figure \ref{fig:fig2a}). The overall amplitude is slightly higher for the northern hemisphere. Moreover, this latitude range $(15-30^{\circ})$ is the transition region from disk dominated to high latitude diffuse halo dominated region, and our result shows that in the transition region the angular power spectra values are considerably different in Northern and Southern hemispheres. This may be due to the complex structure of disk contributing asymmetrically, or variation of structures due to disk halo interaction in the two hemispheres leading to asymmetric structures in density and magnetic fields \citep{simard80,mao12}. We also see that the median $C_{\ell}$ are mostly flat beyond $\ell \geq 1500$ in different latitude bins. In the left panel, we detect a higher residual power at the Galactic plane mostly due to a combination of higher rms. noise (Figure 8 in \citealt{intema17}) and residual point sources. The flattening of the $C_{\ell}$ around the same angular scales ($\sim 0.12^{\circ}$) for all latitude ranges seems to suggest the relative contribution of the increase in rms noise and the DGSE is similar as we move away from the Galactic plane.

We also compare the median values of the measured $C_{\ell}$ in the North Polar Spur (NPS) ($20<{\it l}<40$ and $20<{\it b}<70$) and the southern hemisphere $({\it b}<-20)$. We detect a factor of two increase of the median values for NPS across the entire angular scales. In  Figure \ref{fig:fig4}, we see the same feature where the magnitude of the rms brightness temperature at NPS is large as compared with the southern hemisphere.

\section{Comparison with Single Dish Measurements}
\label{compare}
In this section, we compare our $\delta T_b$ maps with single dish all-sky surveys. The aim is to quantify how much correlation is present between these two maps; usually the single-dish maps are more sensitive to large scale diffuse emission in the sky. Hence, the cross-correlation between the interferometric and signal dish rms. maps will inform us how sensitive the TGSS observations are towards detecting the DGSE. 

%{\bf We use the LFmap \citep{lfmap} to characterize the brightness temperature distribution $(T_b)$ at 150 {\rm MHz}. The LFmap scales the $408$ {\rm MHz} all-sky Haslam map \citep{haslam82} to lower frequencies taking into account three components to the emission: the CMB, isotropic emission from unresolved extragalactic sources, and anisotropic Galactic emission. The spectral index map is derived from \citet{platania03} up to intermediate frequency near $200$ {\rm MHz} and then the 22 {\rm   MHz} map of \citet{roger99} is used to take care of the flattening of the Galactic spectrum below about $200$ {\rm MHz}. Here, we fixed the frequency at 150 {\rm MHz} in LFmap and the resulting map for the all-sky $T_b$ is shown in Figure \ref{fig:fig4}.} The resolution was kept similar to that of the Haslam map ($\sim0.8^{\circ}$). We noticed, as expected, the amplitude of the brightness temperature ($T_b$) is much brighter aaround the Galactic plane. This is mainly due to the contribution from the large scale Galactic Synchrotron emission which dominates at this region.  Also, the North Galactic Spur is clearly visible in this figure.

We use publicly available improved all-sky $408~{\rm MHz}$ Haslam map\footnote{\verb|http://www.jb.man.ac.uk/research/cosmos/haslam_map/|} from \citet{remaz15} with an angular resolution $\sim7$ arcmin which is relevant for studying the foreground contribution in 21-cm signal from the EoR. We downgrade the all-sky map to an angular resolution of $13.7$ arcmin. This is solely done for ease of computation without losing too much information for our purpose.

\begin{figure}
\begin{center}
\includegraphics[width=80mm,angle=0]{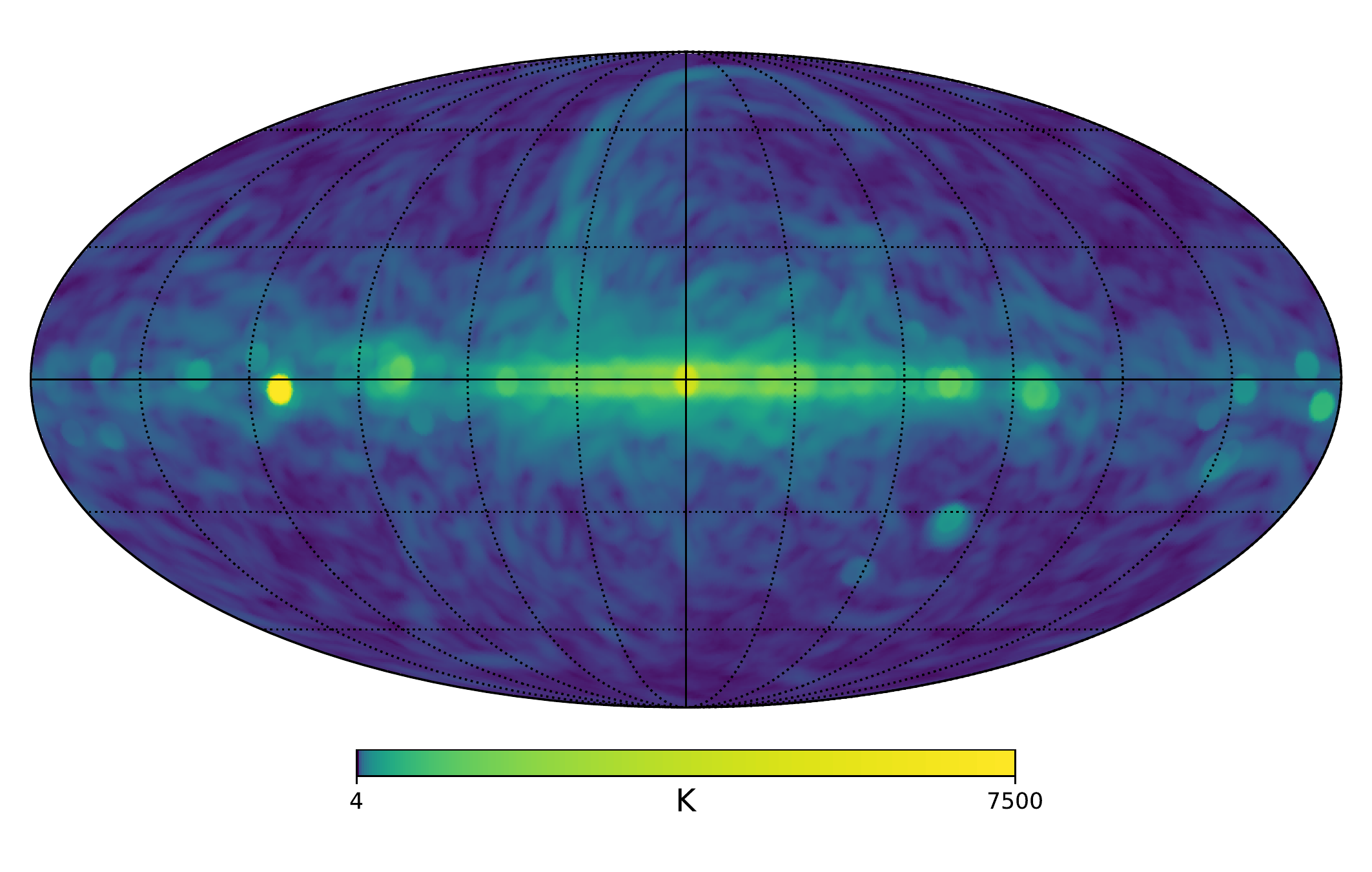}
\put(-235,78){\scriptsize 180}
\put(-120,140){\scriptsize +90}
\put(-120,22){\scriptsize -90}
\put(-71,76){\scriptsize 270}
%\put(-100,140){All Sky Map at $150 {\rm MHz}$}
\caption{Map of the brightness temperature rms of the DGSE at $150 {\rm MHz}$ from an improved all-sky Haslam map. The angular resolution of this map is $13.7$ arcmin which is downgraded from the original $1.7$ arcmin $408~{\rm MHz}$ map \citep{remaz15}. We use an average spectral index $2.695$ from \citet{platania03} to scales at $150~{\rm MHz}$ which is relevant for our study. The rms is calculated within a radius $3^{\circ}$ close to the field of view of GMRT at this frequency.}
\label{fig:fig4}
\end{center}
\end{figure}
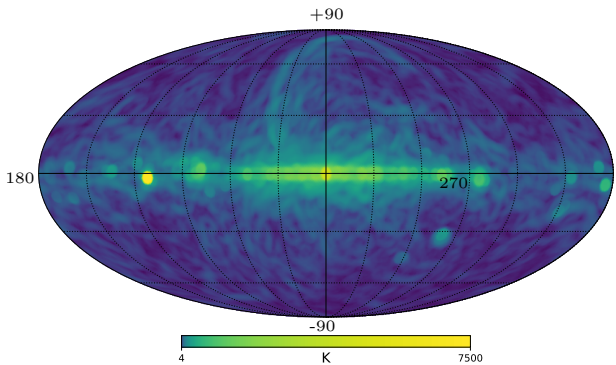

We scale the Haslam map to a lower frequency at $150~{\rm MHz}$ from $408~{\rm MHz}$ using an average spectral index of 2.695, which is typical for DGSE \citep{platania03}. We note here, as the single dish measures the brightness temperature ($T_{b}$) of the sky, we calculate the rms. fluctuations of these maps within a radius of $3^{\circ}$, close to the field of view of GMRT antenna element at $150~{\rm MHz}$. These rms. maps will then give us an equivalent representation of the interferometric observations which are sensitive to the brightness fluctuations of the temperature maps. The rms map with an angular resolution $13.7$ arcmin is shown in Figure \ref{fig:fig4}. We have used this map to cross-correlate with the derived brightness temperature fluctuations from the TGSS measurements.

\begin{figure*}
\begin{center}
\includegraphics[width=180mm,angle=0]{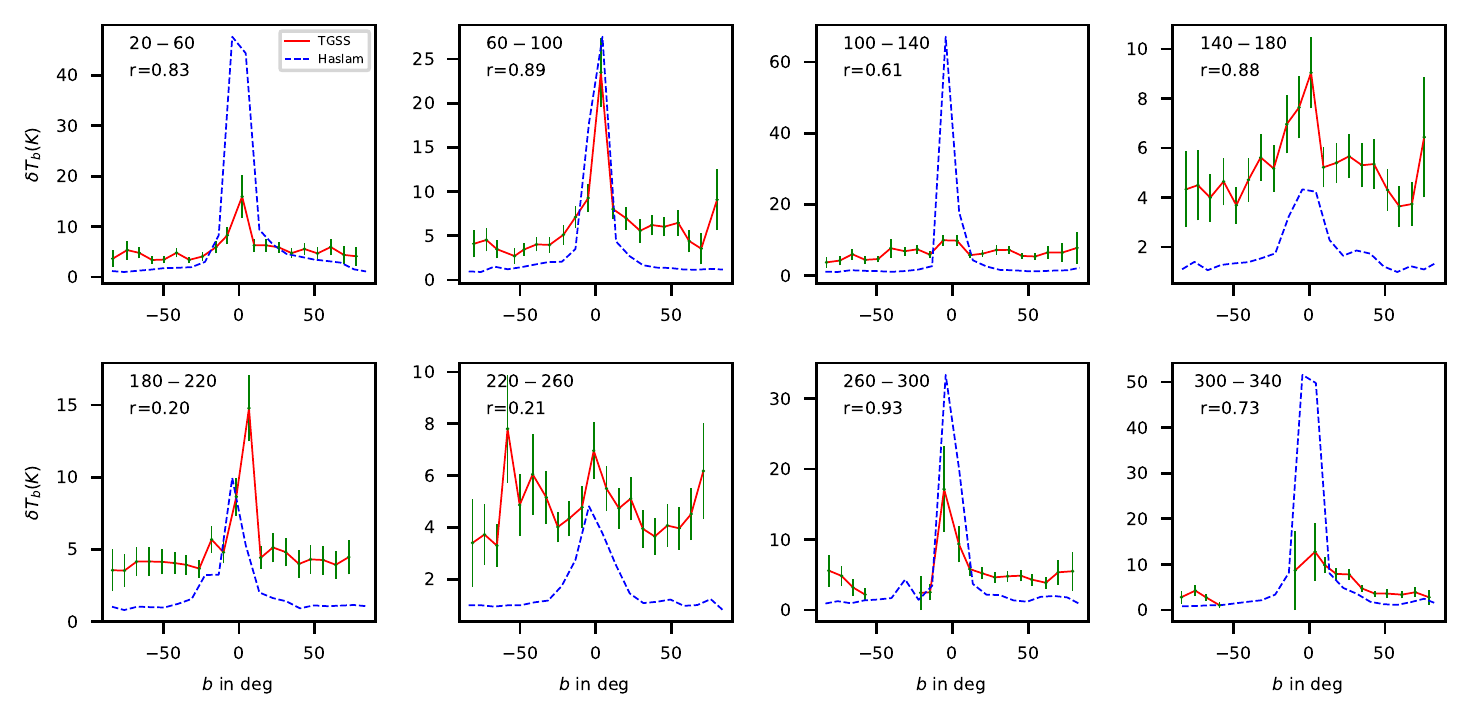}
\caption{The brightness temperature fluctuations from the Haslam and the TGSS survey as a function of galactic latitude for different longitude range mentioned in each panel. The value of the angular multipole used for the TGSS is ${\ell=246}$ which is free from the convolution of the tapering window and primary beam, and likely to be dominated by the DGSE. Here, the blue dashed lines show the rms of the Haslam map divided by $10$, whereas the red solid lines present the $\delta T_b$ values with $1-\sigma$ error bars from the TGSS survey. The corresponding Pearson product-moment correlation coefficients $(r)$ is also shown in each panel.}
\label{fig:fig5}
\end{center}
\end{figure*}

Next, we investigate the correlation coefficient between the TGSS and the Haslam scaled map at $150~{\rm MHz}$ at different longitude and latitude ranges.  For TGSS survey, we use the map at a multipole $\ell=246$ or equivalently $\theta\sim0.73^{\circ}$. As shown in Figure 3 \citep{samir17a}, the residual map at $\ell=246$ is free from the convolution of the
tapering window and primary beam, and likely to be dominated by the
DGSE. Figure \ref{fig:fig5} shows the variation of the brightness temperature fluctuations as a function of galactic latitude for different longitude ranges. Here, we divide the longitude range with an interval of $40^{\circ}$ and show in differnt panels. In Figure \ref{fig:fig5} the blue dashed line presents the rms fluctuations from the Haslam map, whereas the red solid lines show the $\delta T_b$ with $1-\sigma$ error bar from the TGSS survey. Note, we divide the rms. of the Haslam map by a factor of $10$ so that we get a better visualization of the trends of the cross-correlation in Figure \ref{fig:fig5}. We find that for almost all cases, the trend of variation for the TGSS and Haslam map as a function of galactic latitude is quite similar, these curves peak around the Galactic plane and then slowly falls off for higher galactic latitudes. We also noticed some additional peaks in the TGSS measurements (e.g. at $({\it l,b})=(220-260,-60))$, for which the exact reason currently unknown to us.

To quantify the correlation between the Haslam map and the TGSS survey we computed the Pearson product-moment correlation coefficients, defined as, $r_{ij}=\frac{C_{ij}}{\sqrt{C_{ii}*C_{jj}}}$ where $C_{ij}$ is the covariance of $x_i$ and $x_j$ and the element $C_{ii}$ is the variance of $x_i$, 
here $x_i$ and $x_j$ corresponds to the rms of the Haslam map and $\delta T_b$ from TGSS. We present the variation of the correlation coefficient, $r$, for different Galactic longitude ranges in Figure \ref{fig:fig5}. We find $r \geq 0.5$ for most of the longitude ranges ($\sim > 75 \%$). The relatively higher Pearson product-moment correlation coefficient assures us that at this angular scale ($\theta\sim0.73^{\circ}$) of TGSS survey, we are quite sensitive to large scale diffuse emission of the Galactic synchrotron emission.

\section{Summary and Conclusions}
\label{summ}
In this paper, we have estimated the all-sky angular power spectrum of the temperature fluctuations using 150 {\rm MHz} TGSS survey. The angular resolution for this survey is $25\times25$ {\rm arcsec}. The frequency and angular resolution of this survey are relevant for studying the Galactic synchrotron emission, which is one of the main foreground components for detecting the cosmological 21-cm signal from the EoR. 

We present the angular power spectrum, $C_{\ell}$ measurements of the TGSS survey both before and after subtracting the point sources from the data. We find that the measured $C_{\ell}$ before point source subtraction is nearly flat, in the range $10^4-10^5~{\rm mK^2}$, across the measured angular multipoles. This is mainly due to the discrete radio sources which are distributed isotropically all over the sky. The amplitude of the $C_{\ell}$ falls significantly after subtracting the point sources, and we observe that the amplitude is slightly higher at Galactic plane in the angular scale range $0.3^{\circ}$ to $0.8^{\circ}$. We expect that the residual $C_{\ell}$ is likely to be dominated by the Galactic synchrotron emission in these angular scales. However, in the Galactic plane, there will also be an additional contribution from the thermal emission from HII regions, non-thermal emission from supernova remnants and the diffuse synchrotron emission. On the other hand, the measured $C_{\ell}$ at small angular scale (large $\ell$) will be dominated by the unsubtracted point sources in the residual data. We find the resultant all-sky map around a high angular multipole ($\ell$ = 3459 or $\theta \sim 0.052^{\circ}$) is almost isotropic and the derived brightness temperature varies in the range $50-100~{\rm K}$ in different directions.

Looking into the measured $C_{\ell=384}$ across different Galactic latitude range we noticed the $C_{\ell}$ is mostly peaked around the median values for all latitude ranges except at the lower galactic latitude. At lower latitude, the distribution is slightly bi-modal, which can be due to the artefacts of the complex extended sources present in the Galactic plane. We find that the DGSE remains significant over a certain range of multipole on the residual data and the median values from different latitude ranges fall as we move beyond the Galactic plane. The median values of $C_{\ell}$ due to DGSE saturates beyond ${\it b}>30^{\circ}$ and its amplitude becomes much weaker compared to that on the Galactic plane.

We investigated the north and south asymmetry using the residual data for different latitude range. We found that the median $C_{\ell}$ as a function of $\ell$ is almost symmetric for both hemispheres except in the latitude range $15-30^{\circ}$. This latitude range is the transition region from the disk dominated to high latitude diffuse halo dominated region. This may be due to the complex structure of disk contributing asymmetrically, or variation of the structure due to disk halo interaction in the two hemispheres leading to asymmetric structures in density and magnetic field. We also found the $C_{\ell}$ measurement in the NPS is almost a factor of two higher compared to the southern region of the sky.

Cross correlating the Haslam and TGSS brightness temperature fluctuations, we detected a correlation coefficient of $r > 0.5$, which suggests at this angular scales $(0.3^{\circ}$ to $0.8^{\circ})$ we are sensitive to large scale diffuse Galactic Synchrotron emission.

Finally, we plan to undertake a detailed all-sky study of the residual $C_{\ell}$ as a function of the angular multipole. This will be a part of a separate upcoming paper. We expect the measured $C_{\ell}$ to behave as a power-law at low angular multipoles ($\ell \le 550$) and we plan to find out the power-law index from the TGSS survey. This will also enable us to study the variation of the power-law index over different Galactic latitude ranges. This, in turn, can be used as a model for the DGSE for EoR studies, further to study the magnetic field fluctuations and the ratio of random to ordered magnetic fields in the Galactic plane.

We note recently \citet{dolfi19} and \citet{Tiwari19} have used TGSS-ADR1 data sets to calculate the clustering properties of radio sources on very large angular scales $(2<\ell<30)$ and estimated the angular power spectrum from number count statistics. They found the amplitude of the TGSS angular power spectrum is significantly larger than that of the NVSS, which can not be explained by any physically motivated models. The authors indicated some unknown systematic errors are present in the TGSS-ADR1 dataset. Although we are not sensitive to such small angular multipoles using our visibility based estimators, our results also may be influenced by some systematic flux calibration errors ($\sim 10 \%$). There also maybe some other issues like, calibration errors, ionospheric distortion, de-convolution errors during imaging and the point source subtraction, which are significantly more important at low-frequency radio observations. We plan to address these effects with the new release of TGSS-ADR2 data.

\section{Acknowledgements}
We thank the anonymous referee and the scientific Editor for their useful comments and suggestions. SC acknowledge NCRA-TIFR for providing financial support. AG would like to thank the SARAO for support through SKA postdoctoral fellowship, 2016. Some of the results in this paper have been derived using the healpy and HEALPix package. We would like to acknowledge Mathieu Remazeilles for pointing us to the Haslam map (\url{http://www.jb.man.ac.uk/research/cosmos/haslam_map/}). We thank Arianna Dolfi for sharing the TGSS angular two-point correlation function data with us. We thank the staff of the GMRT that made these observations possible. GMRT is run by the National Centre for Radio Astrophysics of the Tata Institute of Fundamental Research.

\end{document}